\documentclass[aps,prx,twocolumn,unsortedaddress,floatfix,nofootinbib,superscriptaddress] {revtex4-2}
\usepackage{amsmath}
\usepackage{amssymb}
\usepackage{bbm}
\usepackage{mathtools}
\usepackage{siunitx}
\usepackage{physics}
\usepackage{graphicx}
\usepackage{dcolumn}
\usepackage{bm}
\usepackage{qcircuit}
\usepackage{svg}
\usepackage[colorlinks]{hyperref}

\newcommand{\bseq}{\begin{subequations}}
\newcommand{\eseq}{\end{subequations}}
\newcommand{\bigO}{\mathcal{O}}

\newcommand{\Norb}{N_\text{orb}}
\newcommand{\bsplit}{\begin{split}}
\newcommand{\esplit}{\end{split}}

\let\oldaddcontentsline\addcontentsline
\renewcommand{\addcontentsline}[3]{}

\definecolor{linkcolor}{RGB}{0,83,166}
\hypersetup{
  colorlinks = true,
  allcolors = {linkcolor}
}

\begin{document}

\newcommand{\mytitle}{Measuring higher-order bulk topology in a programmable quantum simulator}

\title{\mytitle}

\newcommand{\dwave}{\affiliation{D-Wave Quantum, 3033 Beta Ave., Burnaby, BC V5G 4M9, Canada}}
\newcommand{\caltech}{\affiliation{Department of Physics and Institute for Quantum Information and Matter, California Institute of Technology, Pasadena, CA 91125, USA}}
\newcommand{\sfu}{\affiliation{
Department of Physics, Simon Fraser University, Burnaby, BC V5A 1S6, Canada}}
\newcommand{\nmc}{\affiliation{New Mexico Consortium, Los Alamos, NM 87544, USA}}
\newcommand{\ubc}{\affiliation{Department of Physics and Astronomy, and Quantum Matter Institute, University of British Columbia, Vancouver, BC, Canada V6T 1Z1}}

\author{Pratik Sathe}
\email{psathe@dwavesys.com}
\dwave
\nmc

\author{Niclas Heinsdorf}
\caltech

\author{Majid Kheirkhah}
\dwave

\author{Rahul Deshpande}
\dwave 

\author{Fabio Altomare}
\dwave 

\author{Andrew J. Berkley}
\dwave

\author{Martin Dehn}
\dwave

\author{Richard Harris}
\dwave

\author{Emile Hoskinson}
\dwave

\author{Stefan Paul}
\dwave

\author{Gabriel Poulin-Lamarre}
\dwave

\author{Jack Raymond}
\dwave

\author{Chris Rich}
\dwave

\author{Mohammad H. Amin}
\dwave 
\sfu

\author{Marcel Franz}
\ubc

\author{Andrew D.~King}
\email{aking@dwavesys.com}
\dwave

\date{\today}

\begin{abstract}
Topological insulators lie outside the standard Landau paradigm: Instead of being characterized by local order parameters and spontaneous symmetry breaking, their defining properties are encoded in the global structure of their quantum states and manifested as robust boundary states.  Crystalline higher-order topological insulators can extend this principle by hosting localized corner modes in two and higher dimensions. Although these boundary modes can be detected spectroscopically through the local density of states, determining the underlying bulk topology, for example through nested Wilson loops, remains an outstanding experimental challenge. Here we use up to 728 qubits in a superconducting analogue-digital quantum simulator to realize and probe a hierarchy of one-, two-, and three-dimensional topological insulators governed by effective tight-binding Hamiltonians. By reconstructing complex propagators from real-space measurements, we obtain complementary boundary and bulk characterizations: local densities of states reveal localized boundary modes, whereas experimentally reconstructed band projectors yield bulk topological invariants, culminating in a three-dimensional octupole moment extracted through third nested Wilson loops. These results establish a route to realizing, characterizing, and manipulating topologically protected states in programmable three-dimensional quantum matter.
\end{abstract}

\maketitle 
A central property of topological matter is the bulk-boundary correspondence: the topological properties of a material's bulk dictate the nature of states at its boundary~\cite{hasan_colloquium_2010,QiTopological2011,WenColloquium2017}. In conventional, or first-order, topological  insulators, these states occupy codimension-one boundaries, such as edges in two dimensions and surfaces in three dimensions.
The one-dimensional Su-Schrieffer-Heeger (SSH) chain---a model of polyacetylene---provides a minimal example~\cite{SuSolitons1979}: its two dimerizations are distinguished by a quantized polarization and its topological phase supports localized end modes associated with fractional boundary charge~\cite{JackiwSolitons1976a,asboth2016short,Mudry_Chamon_2025}. 

Crystalline symmetries can quantize electric multipole moments, giving rise to higher-order topological phases such as quadrupole and octupole insulators, as exemplified by Benalcazar-Bernevig-Hughes (BBH) models~\cite{BenalcazarQuantized2017,BenalcazarElectric2017}.  These systems extend the bulk-boundary hierarchy to boundaries of codimension greater than one~\cite{SchindlerHigherorder,XieHigherorder2021}: the two-dimensional BBH quadrupole insulator hosts corner modes while its bulk and edges remain gapped, and the three-dimensional BBH octupole insulator exhibits a hierarchy of gapped surfaces and hinges terminating in corner modes. Together with SSH chains, these models provide a natural hierarchy of topological insulators, spanning spatial dimension and boundary codimension from one to three, with charge concentrating at zero-dimensional corner modes.

The challenge of realizing these canonical models has attracted significant attention. Although they are described by a quantum Hamiltonian, their properties can be emulated, either as classical-wave analogues~\cite{serra-garcia_observation_2018,peterson_quantized_2018,PetersonFractional2020,xue_observation_2020,ni_demonstration_2020,BaoTopoelectrical2019} or in reduced forms using momentum-space or compressed spatial encodings~\cite{NiuSimulation2021,ZhaoExperimental2021,KohRealization2024}. Direct real-space realizations have been demonstrated using up to 36 qubits in up to two dimensions~\cite{katz_floquet_2025,qian_programmable_2025}.

\begin{figure*}
  \includegraphics[width=\linewidth]{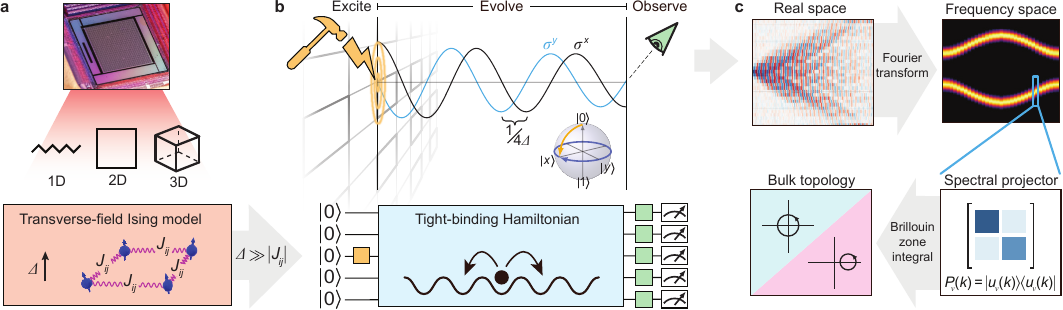}
  \caption{{\bf Realizing and probing bulk topology.} {\bf a}, Programmable two-qubit coupling terms allow our platform to host linear, square, and cubic lattices in the qubits of a superconducting quantum annealing processor, which natively implements a transverse-field Ising Hamiltonian. {\bf b}, In the weak-coupling limit ($\Delta\gg |J_{ij}|$), the low-energy effective Hamiltonian is a noninteracting tight-binding model. Our analogue-digital experimental protocol consists of exciting a single qubit with a rotation $R_y(\theta_e)$, evolving the qubits under a static many-body Hamiltonian for varying evolution time $t$, and measuring the qubits in the $\sigma^x$ or $\sigma^z$ basis; time-delayed $\sigma^x$ measurements approximate $\sigma^y$ measurements. {\bf c}, Site-resolved observations give real-space data. Conversion to frequency space yields a band structure, which allows the calculation of spectral projectors from complex-valued wavefunction measurements via band projector tomography. 
  Computing the associated (nested) Wilson loops enables the extraction of bulk topology.
  }
  \label{fig:1}
\end{figure*}

Here, we realize quantum SSH and BBH models in one, two, and three dimensions, using up to 728 qubits of a superconducting quantum annealing (QA) processor. We use newly developed analogue-digital protocols~\cite{DeshpandeAnalogDigital2026} to access coherent dynamics generated by a static programmable Hamiltonian, marking a significant departure from previous QA applications, which have been confined to coherent global quench dynamics~\cite{king_beyondclassical_2025} or quasi-static measurements~\cite{harris_phase_2018}.

Complementary to the challenge of realizing higher-order topological insulators is the challenge of characterizing them experimentally. This is most often done by measuring model-dependent dynamical signatures~\cite{CardanoDetection2017, MaffeiTopological2018a} or by spectroscopically identifying boundary modes~\cite{serra-garcia_observation_2018,peterson_quantized_2018,PetersonFractional2020,xue_observation_2020,ni_demonstration_2020,BaoTopoelectrical2019,qian_programmable_2025}. Experimentally establishing the bulk-boundary correspondence requires the extraction of bulk topological invariants---in quantized electric multipole insulators, this typically involves a sequence of nested Wilson loops~\cite{BenalcazarQuantized2017},
which requires access to occupied-band Bloch states.
Band topology and geometry have been experimentally extracted in ultracold atomic platforms~\cite{LiBloch2016, FlaschnerExperimental2016,YiExtracting2023};
we follow a more general approach, in which we implement Bloch-state tomography and spectroscopy by experimentally obtaining the complex-valued propagators, leading to an experimental characterization of the Hamiltonian of interest and yielding the desired topological invariants. Together with spectroscopically observed boundary modes and fractionalized spectral weight, we thus probe both sides of the bulk-boundary correspondence.

\section*{Programmable quantum spectroscopy and tomography}

We realize SSH and BBH models using programmable transverse-field Ising model (TFIM) dynamics on a superconducting QA processor (Fig.~\ref{fig:1}a). The processor implements the Hamiltonian
\begin{equation}\label{eq:qa}
    H_\text{TFIM} = -\frac{\Delta}{2}   \sum_{i=1}^N \sigma_i^z + \frac{1}{2} \sum_{i<j} J_{ij} \sigma_i^x \sigma_j^x,
\end{equation}
where the couplings $J_{ij}$ are individually programmable. In the weak-coupling regime ($\Delta{\gg} \abs{J_{ij}}$), excitation-number-changing processes are off-resonant and the dynamics within the single-excitation manifold are governed by the effective tight-binding Hamiltonian
\begin{align}
H_\text{TBM} = \Delta + \sum_{i<j} \frac{J_{ij}}{2} (\ket i \bra j+ \ket j \bra i),\label{eq:eff_H}
\end{align}
where $\ket i \equiv \sigma_i^x \ket{0}^{\otimes N}$ denotes a state with a single spin-flip.

\begin{figure*}\includegraphics[width=\linewidth]{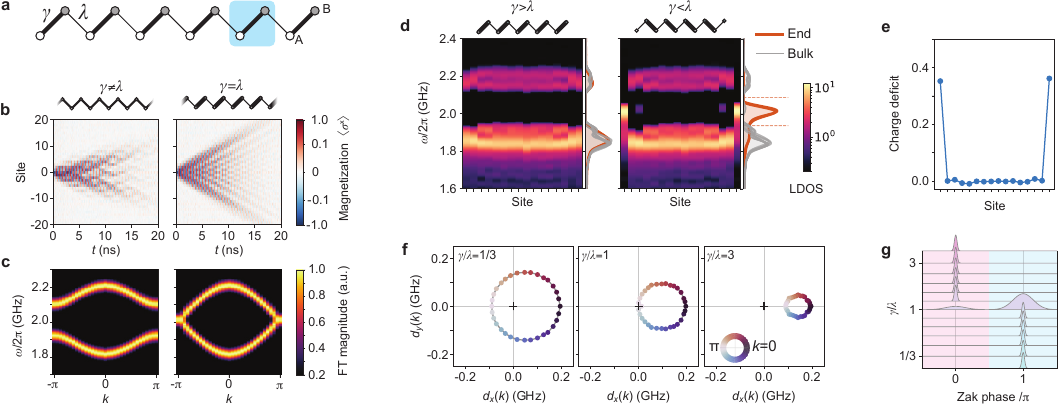}
  \caption{{\bf Dynamics and spectral projector tomography in the SSH model.} {\bf a}, An SSH model consists of $L$ two-orbital unit cells, coupled internally by $\gamma$ and externally by $\lambda$. {\bf b}, Real-space experimental $\sigma^x$ measurements show transport or confinement of a $\tfrac{\pi}{2}$-pulse excitation; $(\gamma,\lambda)$ pairs are, in GHz, $(0.15,0.05)$ and $(0.1,0.1)$, with $\Delta=\SI{2.0}{GHz}$. {\bf c}, Frequency-space $\sigma^x$ measurements, Fourier-transformed from {\bf b}. Data $f(\omega,k)$ are column-normalized and vertically averaged about a central symmetry frequency $\tilde \Delta$ (Methods). {\bf d}, Local density of states on open systems in the trivial (left) and topological (right) phases. {\bf e}, Measuring the topological charge against the trivial background reveals a fractionalized deficit on each end. {\bf f}, Bloch Hamiltonian integral path encircles the origin in the topological phase, resulting in a Zak phase of $\pi$. {\bf g}, The Zak phase jumps from $\pi$ to $0$ mod $2\pi$ as $\gamma{/}\lambda$ is increased through unity. $L{=}32$ for {\bf b}, {\bf c}, {\bf f}, {\bf g} and $L{=}8$ for {\bf d}, {\bf e}.
  }
  \label{fig:2}
\end{figure*}

To probe these effective tight-binding Hamiltonians, we initialize a localized excitation using a single-qubit $R_y(\theta_e)$ rotation, evolve for a variable time $t$, and measure the resulting spin response (Fig.~\ref{fig:1}b) at all sites simultaneously~(ED~Fig.~\ref{fig:ed_protocol}), in the spirit of many-body Ramsey interferometry~\cite{KnapProbing2013}.
The $x$- and $y$-basis measurements with $\theta_e=\tfrac{\pi}{2}$---inferred from time-shifted measurements in a single basis---correspond respectively to the real and imaginary parts of the complex propagator, or equivalently of the single-particle Green's function.
Repeating the protocol over initialization sites and evolution times samples the full real-space propagator, from which we compute frequency-space data and perform Bloch-state tomography~(Fig.~\ref{fig:1}c). We focus on $\theta_e= \tfrac \pi 2$ with $x$-basis measurement but also implement $\theta_e=\pi$ with $z$-basis measurement. Measurements following $R_y(\pi)$ resolve quantum walks, while $R_y(\tfrac{\pi}{2})$ initialization and phase-delayed readout recover the complex propagation amplitudes required for spectral projector tomography.

We first demonstrate the method using the SSH chain, which serves as the simplest topological insulator with a quantized electric moment. We obtain a sequence of progressively stronger signatures of topology: edge-localized dynamics, in-gap end modes, fractionalized boundary spectral weight and, finally, the bulk winding number reconstructed from experimental band projectors.

The SSH model is a chain of $L$ two-site unit cells, with alternating intra-cell and inter-cell hopping terms $\gamma$ and $\lambda$, respectively (Fig.~\ref{fig:2}a). We set $\Delta{=}\SI{2.0}{GHz}$ and vary $\gamma$ and $\lambda$ between $50$ and $\SI{150}{MHz}$. Fig.~\ref{fig:2}b shows qualitatively distinct propagation of a $\tfrac\pi 2$-pulse, measured in the $\sigma^x$-basis, under distinct coupling conditions: uniform ($\gamma{=}\lambda$) and dimerized ($\gamma{\neq} \lambda$) with periodic boundaries, and trivial ($\gamma{>}\lambda$) and topological ($\gamma{<}\lambda$) with open boundaries~(ED~Fig.~\ref{fig:ed_open1d}). We note the clear persistence of an excitation at the boundary in the latter case, suggestive of the expected end mode.

Fourier transforms of the periodic data (Fig.~\ref{fig:2}c) reveal a two-band structure, separated by a gap $\approx 2(\lambda-\gamma)$ at $k=\pm \pi$, closing when $\gamma{=}\lambda$. In open-boundary systems, the extracted bands closely reflect the periodic case except when exciting a qubit at an end of the chain with $\gamma{<}\lambda$, for which the majority of the spectral weight concentrates at the end mode.

Reconstructing local density of states (LDOS) and the bulk invariant requires the complex propagator, including the relative phases of the response on the two sublattices. To reconstruct the wavefunction, we extract $\sigma^x(t)$ and $\sigma^y(t)$ using time shifting~(Fig.~\ref{fig:1}b): In the weak-coupling limit, the qubits precess about the transverse field with frequency $\Delta$, giving the key approximate identity
\begin{align}
\sigma^y_r(t)&\approx \sigma^x_r \left(t+\tfrac{1}{4\Delta} \right).\label{eq:timeshift}
\end{align}
We therefore measure time series of qubit magnetizations $\sigma^x_r(t)$ with time step $\delta t=\tfrac{1}{8\Delta} = \SI{62.5}{ps}$, which allows reconstruction of the site-resolved wavefunction
\begin{align}
\Psi_r(t)= \sigma^x_r(t)+i\sigma^y_r(t).\label{eq:Psi}
\end{align}

Fourier transforming the time variable of the complex return amplitude yields the LDOS, which resolves the energies and spatial support of the single-particle modes. Fig.~\ref{fig:2}d shows the site-resolved densities for topological and trivial cases with open boundaries, with only the topological case harbouring in-gap end modes, consistent with theoretical predictions.

Quantitatively, we infer fractionalization~\cite{JackiwSolitons1976a} by comparing the probability distribution in the topological case against the trivial case for the same value of chemical potential $\mu$~\cite{Mudry_Chamon_2025}.
In particular, we place $\mu$ just above or just below the in-gap end-mode energies and take the difference of the corresponding LDOS integrals (up to $\mu$) in the two cases (Fig.~\ref{fig:2}e).
The difference approaches $\tfrac 12$ at each end but does not reach it, mainly due to broadening of the peaks from the finite time window, relaxation into lower-energy states due to decoherence, and the finite length of the end states (SM). This is indirect evidence of charge fractionalization: a particle deficit in the open topological chain when compared against the trivial open case is equally concentrated, with a fractional weight, at the two ends, due to topological protection~\cite{Mudry_Chamon_2025}.

Although edge confinement, in-gap spectral weight, and fractionalized charge are all consistent with a topological phase, a full characterization requires establishing a non-zero topological invariant, for which we use Bloch-state tomography.

\begin{figure*}
  \includegraphics[width=\linewidth]{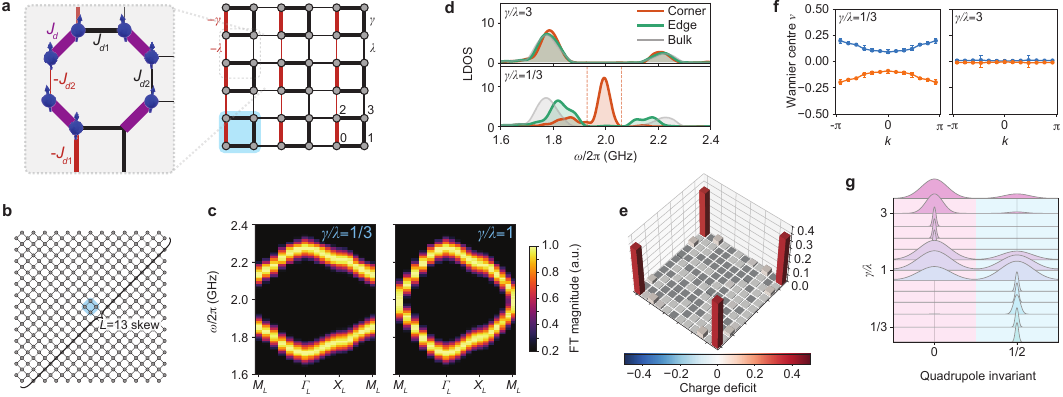}
  \caption{{\bf Quantized electric quadrupole insulator.} 
  {\bf a}, Each site of a logical square lattice is encoded by a strongly coupled two-qubit dimer in a square-octagonal lattice. Inter-dimer couplings generate intracellular and intercellular hopping amplitudes $\gamma$ and $\lambda$, with alternating signs on the vertical bonds, producing a $\pi$ flux through each plaquette.
  {\bf b}, Excitations are initialized in the central four dimers of a skew $L{=}13$ lattice to minimize boundary effects.
  {\bf c}, Experimentally measured bulk spectra along the finite-size high-symmetry path $M_L$-$\Gamma_L$-$X_L$-$M_L$ for two coupling ratios. BBH bands form two twofold-degenerate manifolds whose gap at $M$ closes as $\gamma/\lambda$ approaches unity.
  {\bf d}, Site-resolved LDOS, averaged over site type, for the trivial (top) and quadrupole (bottom) regimes for $L=6$. In the quadrupole regime, the bulk and edge spectra remain gapped and an in-gap corner-localized mode appears.
  {\bf e}, Integrating the difference between the quadrupole and trivial LDOS up to the chemical potential (indicated by the dashed lines in {\bf d}) at the bottom of the band gap yields a spectral-weight deficit localized at the four corners.
  {\bf f}, Wannier spectra reconstructed from the experimental occupied-band projectors. Wilson loops along $x$ and $y$ (equated by graph automorphism) produce gapped Wannier spectra in the quadrupole regime and spectra concentrated near zero in the trivial regime. Error bars represent 95\% bootstrap confidence intervals.
  {\bf g}, Corresponding nested polarizations (bootstrap populations shown). Uncertainties near $\gamma/\lambda{=}1$ and $\gamma{\gg}\lambda$ coincide with near-degeneracy of the energy levels and the Wannier centres, respectively.}
  \label{fig:3}
\end{figure*}

We begin by isolating the experimentally observed bands in frequency and combine measurements from the two source orbitals to reconstruct the momentum-resolved band projectors $\{P_\nu(k)\}$.
To do this, we inject a $\tfrac \pi 2$-pulse excitation, evolve the system and measure $2\sigma^+ = \sigma^x + i\sigma^y$, giving a retarded Green's function.
In the low-temperature and weak-coupling limit, its sublattice-resolved form has a Lehmann representation with residues proportional to the Bloch-band projector matrix elements (SM):
\begin{align}
    \chi^R_{\alpha \beta}(k,\omega) =i \sum_{\nu} \frac{P_\nu^{\alpha \beta}(k)}{\omega - 2\pi E_\nu(k)+i0^+}.
\end{align}
Here, $\alpha$ and $\beta$ are sublattice indices and $\nu$ labels the bands.
By extracting the residue using a method adapted to finite time-grid data, we obtain the $\{ P_n(k)\}$ matrices (SM). 

The full Bloch Hamiltonian $H(k)$ can be constructed from the extracted band energies and the corresponding projectors for any Hamiltonian.
For the SSH model, we construct it from the experimentally extracted values $E_\pm(k)$ and $P_\pm(k)$.
When represented as 
\begin{align}
    H(k) = d_0(k) \sigma_0 + \vec d(k) \cdot \vec \sigma,
\end{align}
the vector $(d_x,d_y)$ traces a circle as $k$ is varied from $0$ to $2\pi$.
This experimentally extracted path encircles the origin for $\gamma {<} \lambda$ but not for $\gamma{>}\lambda$~\cite{asboth2016short}~(Fig.~\ref{fig:2}f), indicating a phase transition between the topological and trivial phases at $\gamma{=}\lambda$~(Fig.~\ref{fig:2}g). The integral over the Brillouin zone of the lower-band Berry connection, which we obtain using the band projector, yields the Zak phase~(Fig.~\ref{fig:2}g).

Together, these analyses provide a complete picture of bulk and boundary topology in a programmable SSH model using site-resolved spectroscopy and Bloch-state tomography.

\begin{figure}
  \includegraphics[width=\linewidth]{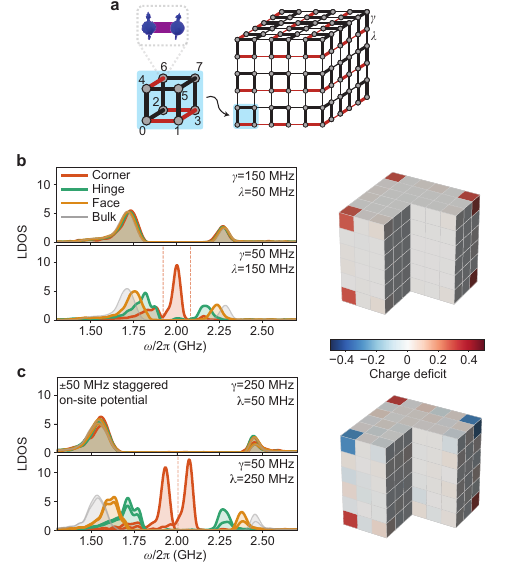}
  \caption{{\bf Octupole boundary phenomena in three dimensions.} {\bf a}, We implement a 3D system using 216 dimers as logical spins ($L{=}3$). {\bf b}, With $\{\gamma,\lambda\}=\{50, 150\}\SI{}{MHz}$, LDOS exhibits gapped bulk, face, and hinge states with topologically protected in-gap corner modes consistent with a quantized electric octupole insulator; fractionalized charge deficit resides primarily on the corner sites. {\bf c}, With $\{\gamma,\lambda\}=\{50, 250\}\SI{}{MHz}$ and a staggered on-site potential $\pm \SI{50}{MHz}$, corner modes bifurcate according to the stagger, remaining in the gap of the lower-codimension states and revealing corner-localized alternate charges indicative of an octupole moment. Cutout sites are removed from the visualization to reveal interior sites.}
  \label{fig:4}
\end{figure}

\section*{Quadrupole and octupole topological insulators}

Having established our approach, we now apply it to realizations of the BBH hierarchy of quantized quadrupole and octupole insulators~\cite{BenalcazarElectric2017}. These models require four-orbital unit cells in two dimensions and eight-orbital unit cells in three dimensions, together with a pattern of signed hopping terms that produces a $\pi$ magnetic flux through each elementary plaquette. 

The available hardware graph does not directly support square and cubic lattices of the required size with the necessary auxiliary state-preparation and measurement qubits. We therefore encode each logical qubit in a strongly coupled two-qubit dimer~(Fig.~\ref{fig:3}a) and treat dimers as effective two-level systems: Tuning the single-qubit transverse field to $\Delta_1{=}\SI{3.0}{GHz}$ and the dimer coupling to $J_d{=}\SI{2.5}{GHz}$, we can treat the dimers as effective Ising moments with a tunnelling rate of $\Delta{=}\SI{2.0}{GHz}$, where inter-dimer couplings $J_{d1}$ and $J_{d2}$ are linearly proportional to the effective hopping terms $\gamma$ and $\lambda$ (SM; ED~Fig.~\ref{fig:ed_dimer}).

We extract the bulk band structure of the two-dimensional model by using a Fourier transform on the propagation of $R_y(\tfrac{\pi}{2})$ excitations initialized in the middle of a 364-dimer lattice with open boundaries (Fig.~\ref{fig:3}b; ED Fig.~\ref{fig:ed_embedding}). The two lower and two upper bands are degenerate. Fixing $\gamma{+}\lambda=\SI{0.2}{GHz}$ and varying $\gamma{-}\lambda$ allows us to tune the band gap observed at the high-symmetry point $M$ ($k_x$, $k_y \in \{\pm \pi\}$)~(Fig.~\ref{fig:3}c).

The LDOS establishes the expected boundary hierarchy: gapped bulk and edge spectra with in-gap corner modes and fractionalized corner spectral weight~(Fig.~\ref{fig:3}d). The charge deficit, extracted as in the 1D case, is consistent with the fractionalized charge approaching $1/2$ on each corner~(Fig.~\ref{fig:3}e). Quantum walks of $R_y(\pi)$ corner excitations are shown in ED~Fig.~\ref{fig:ed_2dquantumwalk}.

A direct topological signature of bulk quadrupole topology can be extracted by taking the projectors onto the energy levels extracted from Fig.~\ref{fig:3}c, computing the Wannier spectrum, and integrating the nested Wilson loop topological invariants---the nested Wilson loop evaluates the Zak phase of the Wannier bands.
As in the SSH case, we determine this invariant directly from the experimentally reconstructed projector onto the occupied bands using spectral projector tomography. For $\gamma{<}\lambda$, we see a gapped Wannier spectrum~(Fig.~\ref{fig:3}f) and the invariants $p_y^{\nu_x}$ and $p_x^{\nu_y}$ are both equal to $1/2$~(Fig.\ref{fig:3}g). 
At the critical point $\gamma{=}\lambda$, we observe a sharp topological phase transition, with the quadrupole moment jumping from $1/2$ to $0$ (modulo $1$) as we increase $\gamma/\lambda$, accompanied by the expected closing of the Wannier gap. This indicates the expected transition between quadrupole and trivial phases.

When all hopping terms have the same sign (i.e., in the 2D SSH model), we resolve a four-band-level spectrum. The LDOS has two gaps, with topologically protected in-gap edge modes in the gaps~(ED~Fig.~\ref{fig:ed_2dssh}).

Finally, we realize the three-dimensional BBH octupole insulator using 432 physical qubits to encode a $3{\times}3{\times}3$ lattice of eight-orbital unit cells~(Fig.~\ref{fig:4}a). With open boundaries, the LDOS (Fig.~\ref{fig:4}b) resolves the complete octupole boundary hierarchy: gapped bulk, face, hinge, and corner states when $\gamma{>}\lambda$, and in-gap corner states when $\gamma{<}\lambda$, with the charge deficit approaching the expected value of $\tfrac 12$ at each corner.

Localization of corner excitations is shown in ED~Fig.~\ref{fig:ed_3dquantumwalk}. To further probe the behaviour of the corner phenomena, we detune the effective transverse field on each site by $\pm \SI{50}{MHz}$, in a bipartite staggered pattern, splitting the degeneracy of the corner modes~\cite{BenalcazarElectric2017}.  This is done with $\{\gamma,\lambda\}=\{50,250\}$~MHz.  When $\gamma{>}\lambda$, the detuning has almost no effect on the LDOS~(Fig.~\ref{fig:4}c). But when $\gamma{<}\lambda$, the symmetry breaking forces a single orientation for the multipole moment, and the corner modes shift up or down following the microscopic staggered pattern. The alternating pattern of positive and negative charges clearly indicates an octupole moment. The analogous phenomenon in the 2D BBH model is shown in ED~Fig.~\ref{fig:ed_2dpolarization}.

\begin{figure}
  \includegraphics[width=\linewidth]{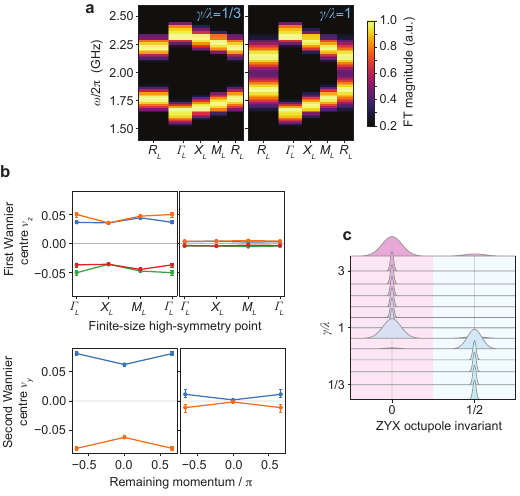}
  \caption{{\bf Quantized electric octupole insulator.}  {\bf a}, Setting a periodic $z$ boundary, we extract a coarse-grained BBH band structure at the finite-size high-symmetry points of the Brillouin zone. {\bf b}, By calculating Wilson loops along $z$ and then along $y$, we obtain the first (top) and second (bottom) nested Wannier spectra for $\gamma/\lambda = 1/3$ (left) and $3$ (right). {\bf c}, ZYX octupole invariant on the third nested Wilson loop; a value of $1/2$ indicates an octupole insulator. Error bars represent 95\% bootstrap confidence intervals across symmetries of the system and ridge lines indicate the bootstrap population.}
  \label{fig:5}
\end{figure}

As in the 2D case, we want to probe the band structure by exciting in the bulk as far from the boundary as possible; with $L{=}3$, this is challenging. Here we apply a periodic $z$ dimension, increasing the number of interior cells and available symmetries. We also reduce the length of our time series from $\SI{10}{ns}$, used in the 2D case, to $\SI{6}{ns}$, to restrict to times where the excitation is primarily away from the boundary, thus minimizing finite-size effects.  We can resolve a coarse BBH band structure~(Fig.~\ref{fig:5}a), enabling band projector tomography.
 
To compute the octupole moment, we first compute the Wannier spectrum from a Wilson loop along a single dimension ($z$, to maximize stability). We then successively evaluate the nested Wilson loops along $y$ and $x$, with the second nesting yielding the octupole moment. Fig.~\ref{fig:5}b shows the two nested Wannier spectra---the first is computed through the 2D high-symmetry points, and the second is computed along a 1D path through the Brillouin zone. As in the lower dimensions, we find a phase transition between the topological and trivial phases as we tune $\gamma/\lambda$ through 1 from below (Fig.~\ref{fig:5}c). 

In the zero-flux 3D SSH model, we see four bulk band levels with a three-band in-gap surface mode~(ED~Fig.~\ref{fig:ed_3dssh}).  

\section*{Outlook}

We have demonstrated programmable quantum realizations of SSH and BBH models---a hierarchy of canonical quantized electric dipole and multipole insulators with dimension and codimension ranging from one to three---using fixed-Hamiltonian dynamics on hundreds of qubits. We use this testbed to experimentally validate higher-order bulk-boundary correspondence in a programmable quantum simulator by combining local spectroscopy and bulk band projector tomography.

The calculation of bulk topological invariants directly from experimental data represents a promising step forward in the self-consistent realization and measurement of programmable quantum matter. It requires high-precision extraction of band energies and complex-valued wavefunction measurements, which we achieve using time-series measurement with fine time resolution.  Extending to three dimensions requires hundreds of lattice sites for the octupole invariants to be resolved; this geometry remains beyond the demonstrated state of the art for modalities such as gate-model superconducting qubits and trapped ions. By computing momentum-resolved spectral projectors onto the extracted energy bands, we obtain the dominant action of the full Hamiltonian, which, in turn, allows direct calculation of topological invariants.  
This complements the standard approach of measuring the density of states with in-gap boundary modes via spectral response.

These experiments remain classically tractable because their dynamics are restricted to the single-excitation sector, whose dimension grows only linearly with lattice size. 
By realizing multiple-excitation sectors in two- and three-dimensional lattices, we could instead access interacting many-body dynamics in a Hilbert space that grows exponentially with system size. This would open a route to interaction-driven topological phases, correlated states in topological flat bands, and nonequilibrium many-body phenomena in programmable quantum systems.

\newpage

\section*{Code and Data Availability}

Experimental results, figures, and analysis code are available at the Zenodo repository \url{https://doi.org/10.5281/zenodo.23061551}.

\section*{Acknowledgements}
The authors are grateful for the D-Wave teams whose work in designing, fabricating, deploying, and calibrating the QPU makes this research possible.  We thank Arnab Banerjee and Christopher Roth for useful discussions.
NH acknowledges funding through the Feodor-Lynen fellowship awarded by the Humboldt Foundation.
MF is supported by NSERC and CIFAR.

\section*{Methods}

These experiments were performed on a D-Wave Advantage2 processor located in Burnaby, BC, Canada.  The general analogue-digital protocol used is shown in Fig.~\ref{fig:ed_dimer}. In the 1D case, multiple disjoint embeddings of the system under study are used in parallel and the statistics are averaged. A QPU call yields 200 shots per embedding per call. A single QPU call has fixed system geometry and couplings ($L$, $\gamma$, $\lambda$, $\Delta$), fixed time $t$, and fixed sign of the polarizing flux bias used to initialize quenched qubits.  Qubit $\sigma^x$ and $\sigma^z$ magnetizations for identical parameterizations other than the polarizing flux bias sign are added with the appropriate signs to cancel out static off-basis biases.  

Frequency-space data, Wannier spectra, and topological invariants are calculated in 1D, 2D, and 3D using time-series lengths of $\SI{20}{ns}$, $\SI{10}{ns}$, and $\SI{6}{ns}$, respectively.  
Throughout, we set the Planck constant $h$ to 1.
Defining $J = \gamma+\lambda = \SI{0.2}{GHz}$, this extends to a normalized time of $Jt/\hbar = 8\pi \approx 25$ in the 1D case, as seen in transport in~Fig.~\ref{fig:2}b, right panel.

\subsection*{Frequency-space analysis}

The sequence of data analysis begins with the calculation of LDOS.
The LDOS is extracted per qubit based on the FT of the phase-sensitive wavefunction extracted from Eq.~\eqref{eq:Psi}, with the initial phase rotated to $0$.  The LDOS is truncated to be non-negative.

Since bulk dynamics are initialized with a finite-time quench of the source qubit, there is a transient crossover between the initial and dynamical regimes.  We therefore choose an optimal point $t{=}0$ using the site-averaged integrated LDOS as the figure of merit.  The integral is taken over the region of interest, e.g.,~1.6--\SI{2.4}{GHz} in Fig.~\ref{fig:2}.  This value of $t{=}0$ is calculated per model and propagated throughout other analyses (SM).

The bipartite structure of the hopping terms in the lattice guarantees (an energy shifted) chiral symmetry implying spectral symmetry about the energy $\tilde\Delta$, which we also obtain from the LDOS analysis.  We take it as the peak location of the corner or end mode for $\gamma/\lambda=1/3$; we use the SSH model in 1D and the BBH model in 2D and 3D.  Thus we obtain three values of $\tilde\Delta$, one for each dimensionality.  In all cases $\tilde\Delta$ is within 1\% of $\Delta=\SI{2}{GHz}$, the expected spectral midpoint from theory.

In all dimensions, we observe an underpopulation of higher frequencies compared to lower frequencies; the same was observed in Ref.~\cite{serra-garcia_observation_2018}, and, in our case, attributable to open quantum system effects and in qualitative agreement with Bloch-Redfield numerical simulations. We therefore calculate charge deficits using the frequency-symmetrized LDOS about the central point $\tilde\Delta$.

Band structures are processed using the available symmetries obtained only from the hopping terms in the lattice.  This includes frequency symmetry about $\tilde\Delta$, symmetries between orbitals, rotational  and reflective symmetries about periodic boundaries (in 1D and 3D), reflective symmetry about open boundaries, and equivalence between $x$ and $y$ dimensions (in 2D and 3D).  These symmetries include those that involve a combination of graph automorphism and gauge transformation.

\subsection*{Computing topological invariants}

Band energies were extracted from frequency-space data by fitting to a sum of two Lorentzians.  These energies were then used to invert the spectral-leakage matrix, yielding the occupied-band projector at each $k$.

In 1D, experiments were repeated for rotations of the chain every 8 sites to suppress the effects of site-dependent device variations.  Since reflecting the periodic lattice about the excitation point has the effect of swapping $\gamma$ and $\lambda$, the cases $\gamma{>}\lambda$ were not programmed separately, but rather extracted from separate analysis of $\gamma{<}\lambda$ data. 

In two and three dimensions, we show frequency-space data across high-frequency cuts.  For finite size, these cuts do not reach the full extent of the Brillouin zone; the sampled Brillouin zone points closest to the high-symmetry points in 2D are
\begin{align*}
\Gamma_L&=(0,0)\\
X_L&=(\tfrac{L-1}{L}\pi,0)\\
M_L&=(\tfrac{L-1}{L}\pi,\tfrac{L-1}{L}\pi).
\end{align*}
In 3D they are
\begin{align*}
\Gamma_L&=(0,0,0)\\
X_L&=(\tfrac{L-1}{L}\pi,0,0)\\
M_L&=(\tfrac{L-1}{L}\pi,\tfrac{L-1}{L}\pi,0)\\
R_L&=(\tfrac{L-1}{L}\pi,\tfrac{L-1}{L}\pi,\tfrac{L-1}{L}\pi),
\end{align*}
leading to the gap at $R_L$ remaining visibly open for $L{=}3$ even when $\gamma{=}\lambda$.

In two dimensions, we used the lattice shown in Fig.~\ref{fig:3}b as embedded in Fig.~\ref{fig:ed_embedding}.  For evaluating Fourier-transformed data, we padded the ``skew'' $L{=}13$ lattice with zero-magnetization sites and evaluated the system as though periodic.  We combined statistics from four central excitation locations.  Bootstrapping was performed over these four sites.

In three dimensions, there are 24 symmetry-related central sites due to the periodic $z$ dimension.  In both 2D and 3D, symmetry-equivalent propagators were generated using lattice automorphisms and their gauge transformations.

The 2D quadrupole and 3D octupole diagnostics were calculated using, respectively, two and three successive Wilson loops~\cite{BenalcazarElectric2017}. These characterize the geometric phase accumulated around a closed Brillouin-zone path.  In 3D, the first loop is taken along the periodic $z$ dimension; the $x$ and $y$ dimensions are related by symmetry.  Further details are given in the Supplementary Material.

Bootstrapping was performed on 200 resamplings of symmetries.

\subsection*{Use of large language models}

Large language models (LLMs) were used to accelerate code development for graph embedding algorithms, experimental scripts, and data analysis.  Initial experiment and analysis code were written by hand and analyses have been independently confirmed.  The output of graph embedding scripts was verified as a straightforward and necessary element of the experimental workflow.

\setcounter{table}{0}
\setcounter{figure}{0}

\renewcommand{\theHfigure}{ED.\arabic{figure}}
\renewcommand{\theHtable}{ED.\arabic{table}}

\renewcommand{\figurename}{Extended Data FIG.}
\renewcommand{\tablename}{Extended Data Table}

\begin{figure*}
  \includegraphics[width=\textwidth]{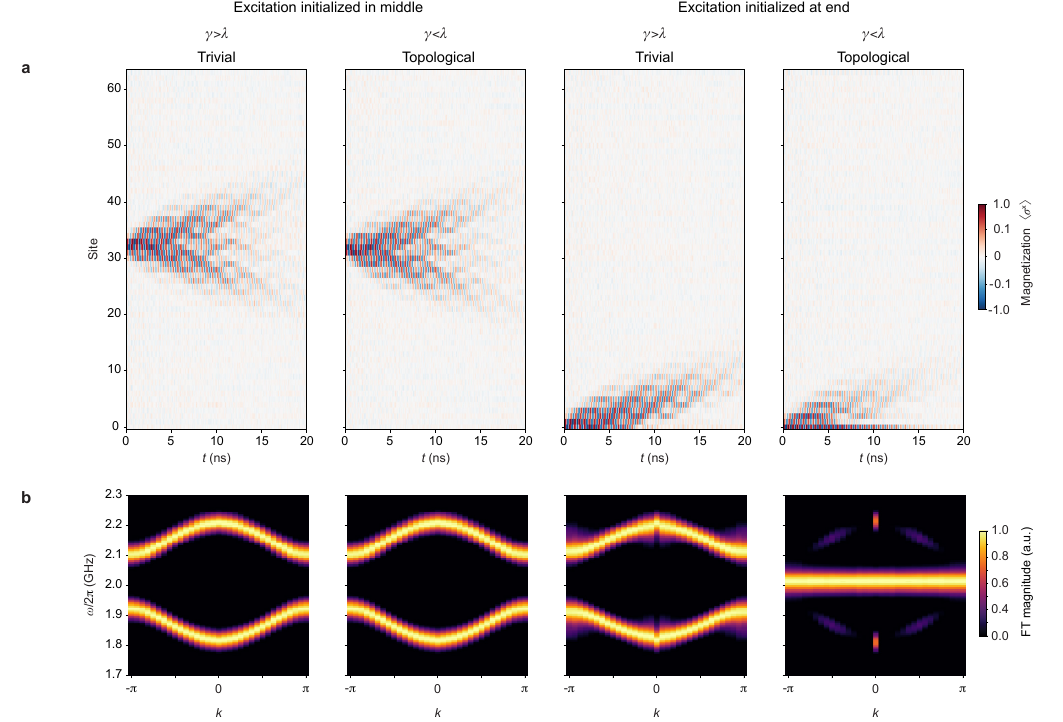}
  \caption{{\bf Open-boundary SSH chains.} {\bf a}, Propagation of a $\tfrac{\pi}{2}$ pulse, corresponding to Fig.~\ref{fig:2}b.  {\bf b}, Fourier-transformed frequency-space data, corresponding to Fig.~\ref{fig:2}c.  First two columns have the excitation initialized on a central site, and second two columns have the excitation initialized on an end site.}
  \label{fig:ed_open1d}
\end{figure*}

\begin{figure*}
  \includegraphics[width=\textwidth]{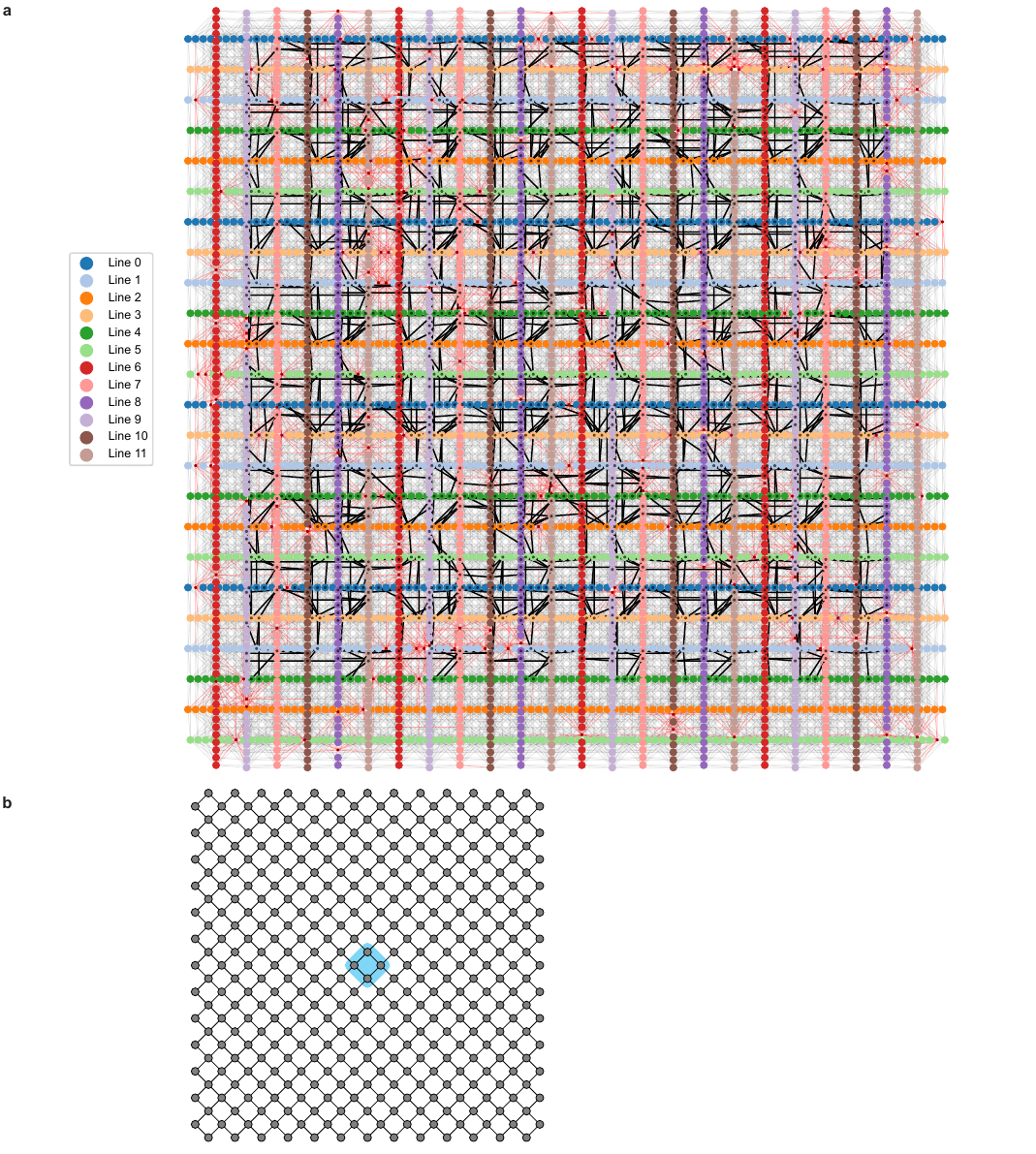}
  \caption{{\bf Embedding of 2D system into qubit graph.}  {\bf a}, 728 qubits are coupled in a square-octagonal layout to represent a 2D square lattice with 364 dimers. 364 ancilla qubits are used for qubit excitation (on lines 8 and 9) and another 364 ancilla qubits are used for measurement (on lines 0 and 4). {\bf b},~Resulting skew square lattice. Excitations are initialized at the four central sites to simulate bulk behaviour in the thermodynamic limit.}
  \label{fig:ed_embedding}
\end{figure*}

\begin{figure*}
  \includegraphics[width=\textwidth]{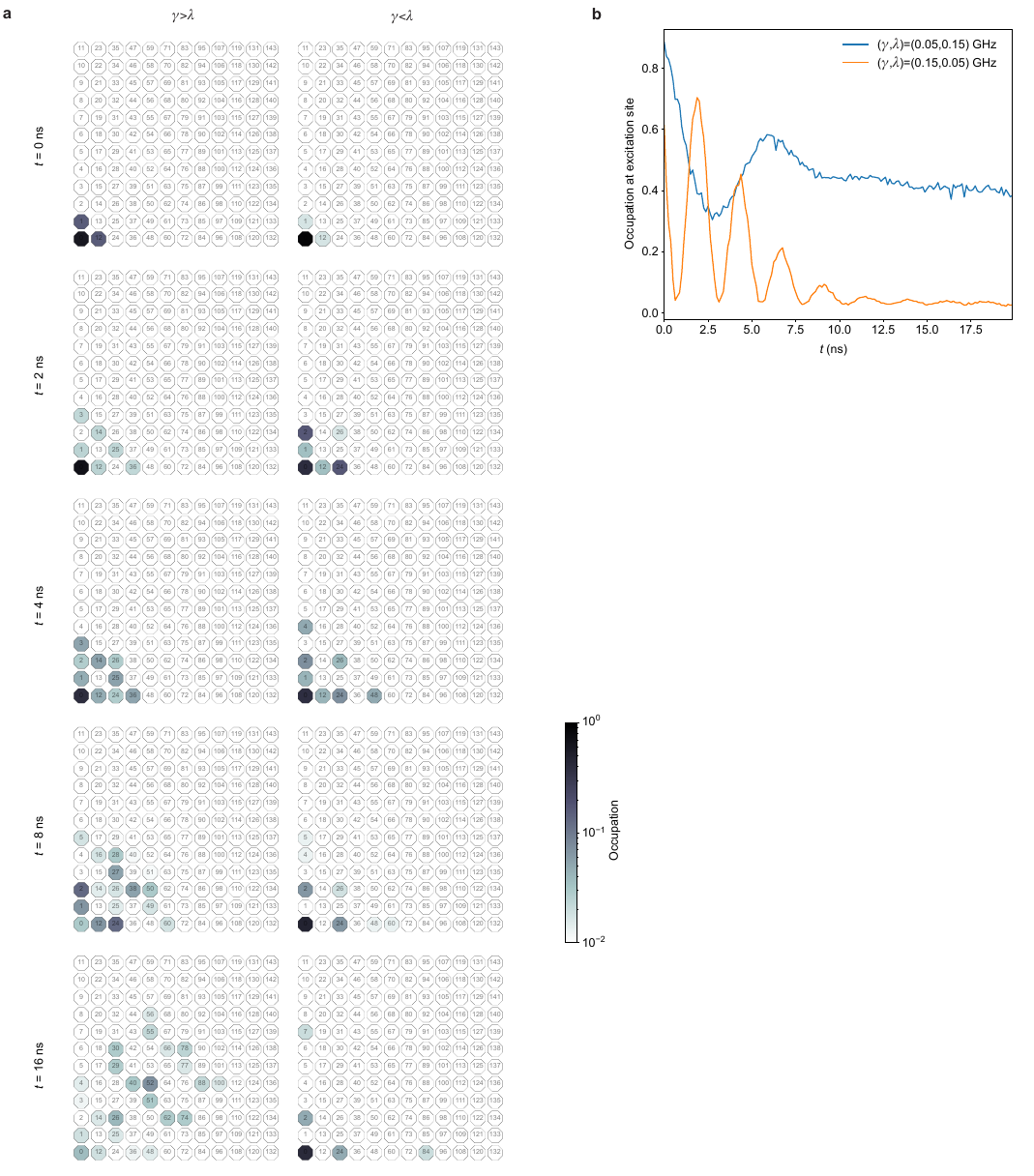}
  \caption{{\bf Quantum walk in a 2D BBH model.}  A $\pi$ pulse is used to initialize an excitation at the lower-left corner and all sites are measured in the $z$ (energy) basis. {\bf a}, Real-space tracking of excitation propagation, with background population $\approx 0.03$ subtracted and total population then normalized to 1. 
  {\bf b}, Initial excitation site occupation for trivial and topological phases, with clear localization in the topological phase.}
  \label{fig:ed_2dquantumwalk}
\end{figure*}

\begin{figure*}
  \includegraphics[width=\textwidth]{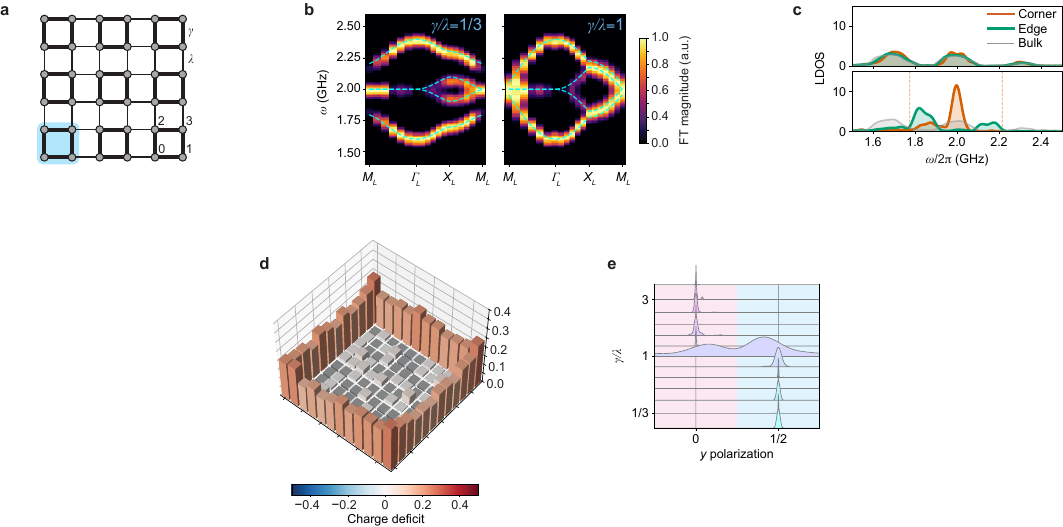}
  \caption{{\bf Two-dimensional SSH model.}
  {\bf a}, Setting all hopping terms to the same sign removes the plaquettes' $\pi$ flux, leaving a two-dimensional SSH model.
  {\bf b}, Fourier-transformed measurements from central excitations ($L{=13}$ skew geometry) show a splitting into four bands, in contrast to the twofold-degenerate ridges of the BBH model.  Theoretical band values are shown in dashed cyan lines.
  {\bf c}, LDOS ($L{=}6$) shows a two-gap structure in the trivial phase, whereas the topological phase ($\gamma{<}\lambda$ has topologically protected in-gap edge states.
  {\bf d}, Integrating up to a chemical potential $\mu$ at the bottom of the lower gap reveals a charge deficit uniformly distributed around the boundary of the lattice.
  {\bf e}, Using $y$-polarization as a topological invariant reveals a transition from a topological phase to a trivial phase near $\gamma{=}\lambda$.  Here, unlike in the BBH models, band projectors are calculated from experimental data using theory ridges instead of experimentally extracted ridges.}
  \label{fig:ed_2dssh}
\end{figure*}

\begin{figure*}
  \includegraphics[width=\textwidth]{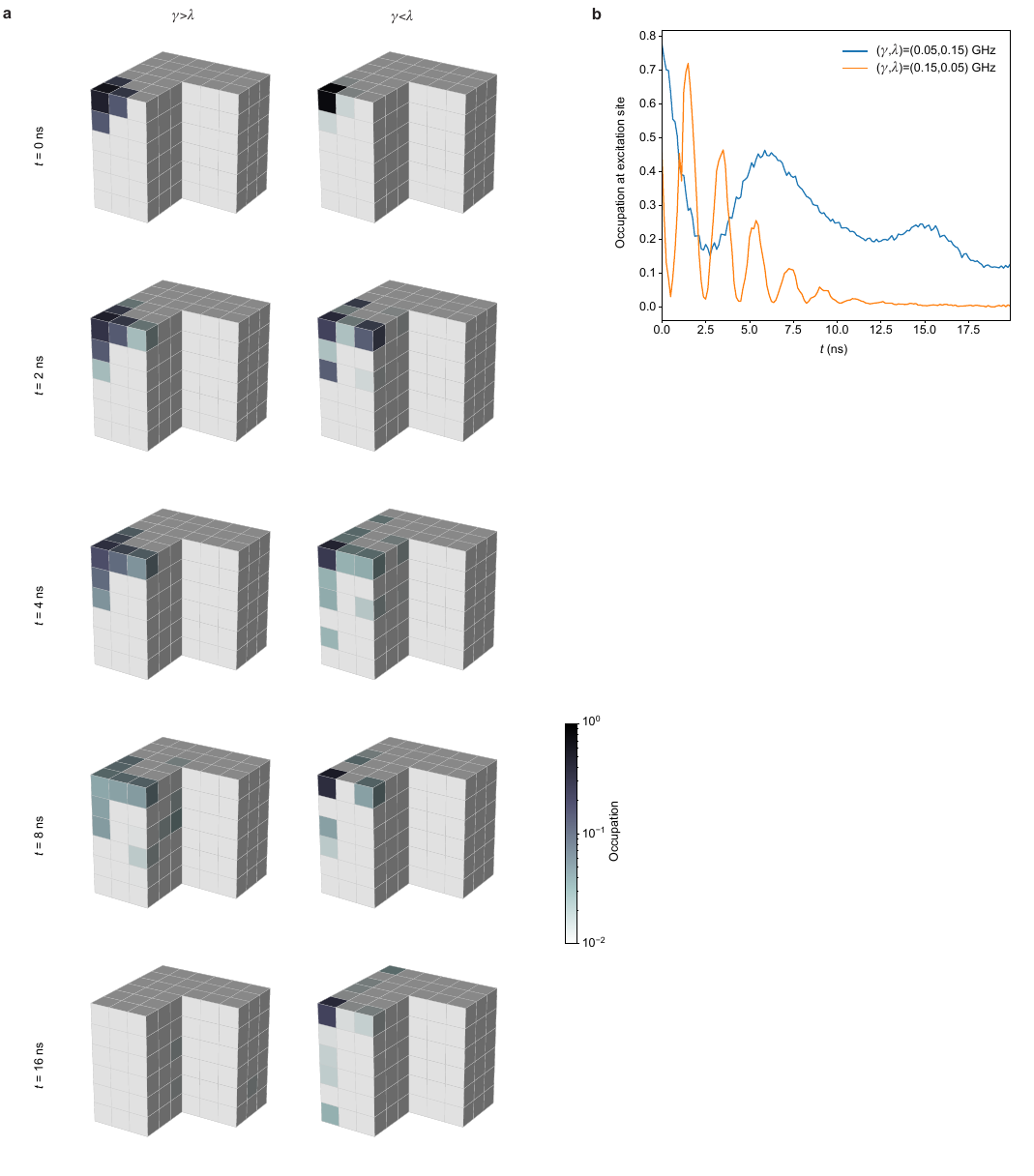}
  \caption{{\bf Quantum walk in a 3D BBH model.}  A $\pi$ pulse is initialized in the corner and all sites are measured in the $z$ (energy) basis. {\bf a}, Real-space tracking of excitation propagation, with background population $\approx 0.03$ subtracted and total population then normalized to 1.   Cutout sites are removed from the visualization to reveal interior sites.
  {\bf b}, Initial excitation site occupation for trivial and topological phases, with clear localization in the topological phase.}
  \label{fig:ed_3dquantumwalk}
\end{figure*}

\begin{figure*}
  \includegraphics[width=\textwidth]{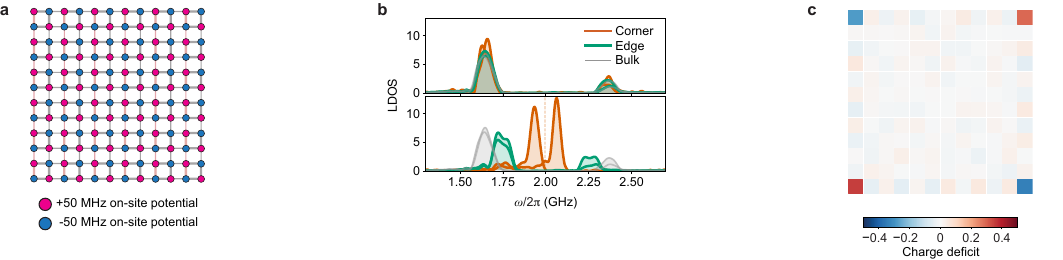}
  \caption{{\bf Polarization in two-dimensional BBH model.} 
  {\bf a}, Analogous to the 3D case in Fig.~\ref{fig:4}, we place a staggered on-site potential of $\pm\SI{50}{MHz}$ on the sites of a 2D BBH lattice with $L{=}12$.  This is achieved at a dimer level by detuning the individual qubit transverse fields by $\pm\SI{54.2}{MHz}$.
  {\bf b}, With $\{\gamma,\lambda\}=\{50,250\}\SI{}{MHz}$, the staggered on-site potential has no significant effect on the site-resolved LDOS when $\gamma{>}\lambda$; shown here are the corner/edge/bulk sites for each stagger bipartition.  When $\gamma{<}\lambda$, the corner modes appear and bifurcate into two classes depending on the bipartition class.
  {\bf c}, Charge deficit obtained by subtracting the topological-phase LDOS from the trivial-phase LDOS and integrating the difference up to the spectral midpoint shows a staggered corner pattern indicative of a quantized quadrupole electric insulator.}
  \label{fig:ed_2dpolarization}
\end{figure*}

\begin{figure*}
  \includegraphics[width=\textwidth]{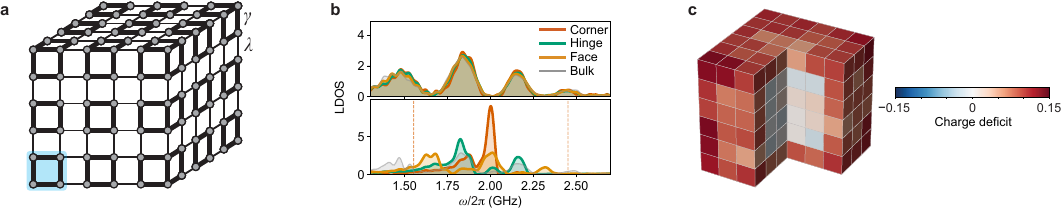}
  \caption{{\bf Three-dimensional SSH model.} {\bf a}, Plaquette flux is zero when all hopping terms have the same sign.
  {\bf b}, In this case, the trivial phase (top) has a four-band structure with the lower gap primarily occupied by facial boundary sites.
  {\bf c}, Excess charge is distributed across the surface of the open-boundary lattice. Cutout sites are removed from the visualization to reveal interior sites.}
  \label{fig:ed_3dssh}
\end{figure*}

\let\addcontentsline\oldaddcontentsline

\clearpage
\onecolumngrid
\begin{center}
\textbf{\large Supplementary Materials:\\ \mytitle}
\end{center}

\tableofcontents
\setcounter{equation}{0}
\setcounter{table}{0}
\setcounter{figure}{0}
\renewcommand{\figurename}{FIG.}
\renewcommand{\tablename}{TABLE}
\renewcommand{\thefigure}{S\arabic{figure}}
\renewcommand{\thetable}{S\arabic{table}}
\renewcommand{\theequation}{S\arabic{equation}}
\renewcommand{\theHfigure}{S\arabic{figure}}
\makeatother

\pagenumbering{arabic}
\setcounter{page}{1}

\section{QPU methods}\label{sec:qpumethods}
Throughout, we set the Planck constant $h$ to $1$.
\subsection{Embedding lattices and auxiliary qubits}

Each many-body lattice of target qubits, whose dynamics we observe using auxiliary source and detector qubits, must be embedded as a subgraph isomorphism into the QPU's qubit connectivity graph. Furthermore, we need certain coverings of the qubits with source and detector qubits, and all of these qubits are additionally constrained to reside on certain control lines: Target, detector, and source qubits follow different Hamiltonian schedules, and must therefore be on separate control lines~\cite{DeshpandeAnalogDigital2026}.

In principle, these embeddings can all be found using a subgraph isomorphism solver such as the Glasgow Subgraph Solver~\cite{mccreesh_glasgow_2020}.  However, this can become computationally onerous for large 2D and 3D lattices, so we also use tailored constraint-based embedding algorithms.

\subsection{Encoding a single logical qubit using multiple physical qubits}
Due to the requirement of reserving source and detector qubits for each target qubit, we are unable to natively embed large square and cubic lattices required to realize the quadrupole and octupole insulator models.
Hence, we realize these models by using dimerization, wherein a pair of strongly coupled physical qubits serves as a single logical qubit; see Fig.~\ref{fig:3}a in the main text.

An individual dimer has a Hamiltonian that may be written as
\begin{align}
    H_\text{dimer} = \tfrac{J_d}{2} X_1 X_2 - \tfrac{\Delta}{2}(Z_1 + Z_2).
\end{align}
We use inter-dimer couplings $J$ that are much smaller than both $\Delta$ and $J_d$, so that the effective dynamics of each dimer is constrained to its eigen-subspace corresponding to the two lower energies. 

The effective Hamiltonian within this subspace is 
\bseq
\begin{align}
    H_\text{eff}^\text{dimer} &= -\tfrac{\Delta_\text{eff}}{2} \sigma_z^\text{eff}, \\
    \text{with } \Delta_\text{eff} &= \operatorname{sgn}(J_d) \left( \tfrac{J_d}{2} - \sqrt{\Delta^2 + \tfrac{J_d^2}{4}} \right).
\end{align}
\eseq
Here, $\sigma_z^\text{eff}$ is a Pauli matrix in an appropriate basis within the lower energy subspace of the dimer. 
(We ignore the constant energy offset term that also appears in the expression.)

An Ising interaction between two logical qubits is encoded in terms of an Ising interaction of strength $J$ between two qubits belonging to adjacent dimers.
The effective Ising interaction between two logical qubits $A$ and $B$ can be expressed as
\bseq
\begin{align}
    H_\text{eff}^\text{(dimer A)-(dimer B)} &= \tfrac{J_\text{eff}}{2} \sigma_x^A \sigma_x^\text{B} \\
    \text{with } J_\text{eff} &= -\tfrac{J}{2} \left[ 1 + \frac{\abs{J_d}/2}{\sqrt{\Delta^2 + {J}_d}^2 /4}\right].
\end{align}
\eseq
In our experiments, the intra-dimer coupling $J_d$, and inter-dimer couplings $J$ are chosen to yield the desired logical transverse-field strength $\Delta_\text{eff}=\SI{2.0}{GHz}$ and Ising interacting strengths $\mathcal J_\text{eff}$.

\begin{figure*}
  \includegraphics[width=\textwidth]{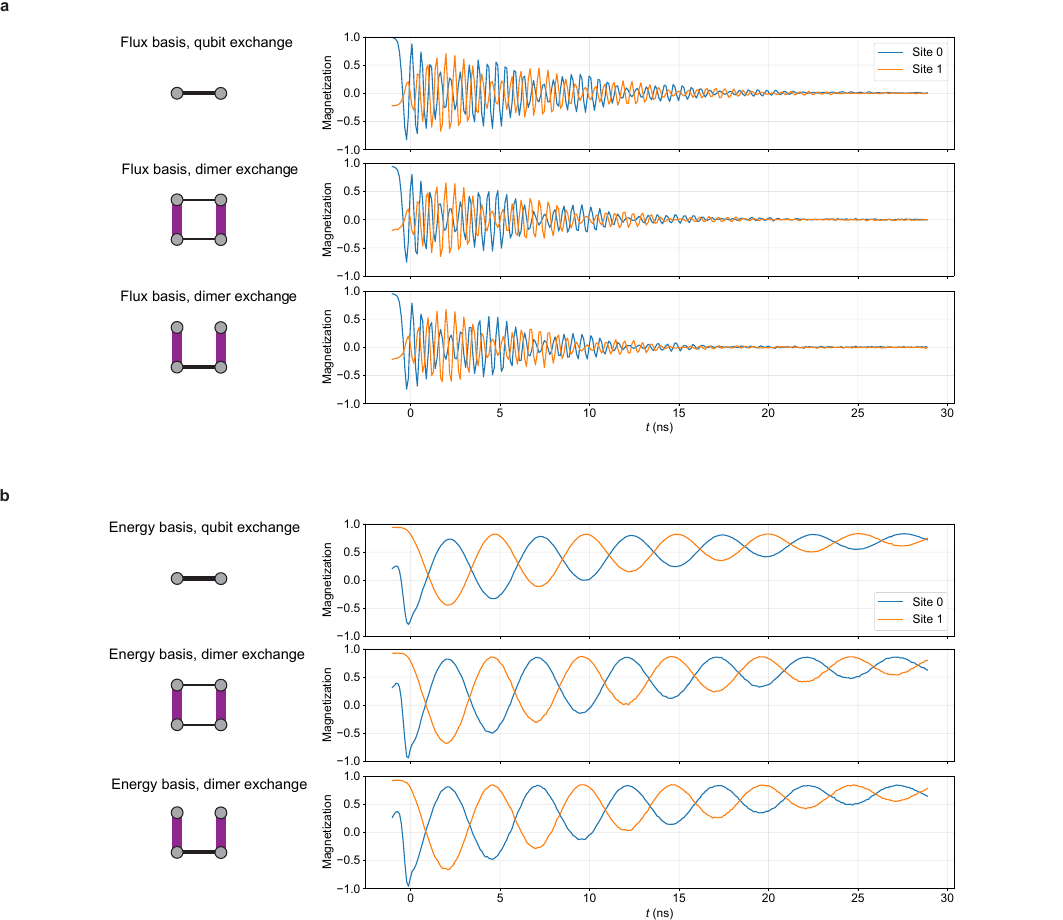}
  \caption{{\bf Dimers as effective logical qubits.}  We measure the exchange between two coupled qubits or two coupled dimers in the flux basis ({\bf a}, $\tfrac{\pi}{2}$-pulse, measure in $\sigma^x$ basis) and the energy basis ({\bf b}, $\pi$-pulse, measure in $\sigma^z$ basis).  Two distinct dimer coupling schemes are shown: one in which the exchange energy is split between two couplers, and one in which only one coupler joins the dimers.  In agreement with theory, setting $\Delta_d=\SI{3.0}{GHz}$ and $J_d=\SI{2.5}{GHz}$ provides close agreement with physical qubits at $\Delta=\SI{2.0}{GHz}$; to achieve comparable exchange frequencies, inter-dimer couplings are set $13/9$ stronger than the couplings between single qubits.}
  \label{fig:ed_dimer}
\end{figure*}

To validate the ability of dimers to emulate single degrees of freedom, we track exchange dynamics of $R_y(\tfrac{\pi}{2})$ and $R_y(\pi)$ excitations in the $x$ and $z$ bases, respectively, observing close agreement~(Fig.~\ref{fig:ed_dimer}).

\subsection{Analogue-digital protocols} \label{subsec:ad_protocols}

\begin{figure*}
  \includegraphics[width=\textwidth]{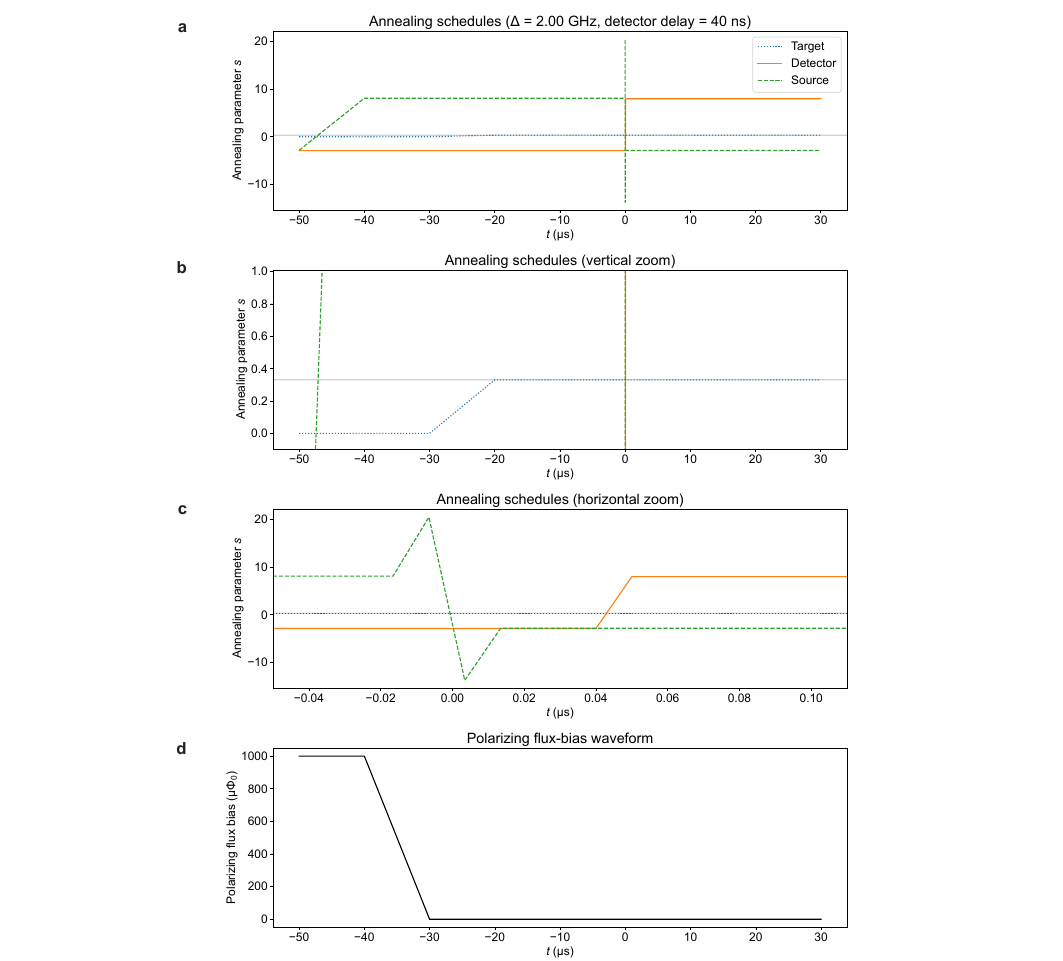}
  \caption{{\bf Analogue-digital control protocol.} {\bf a}, Target, source, and detector qubits, on distinct sets of annealing lines, follow different annealing schedules.  {\bf b--c}, Vertical and horizontal details show the regions of interest ($0\leq s \leq 1$ and $t\approx 0$).  {\bf d}, State preparation is achieved by polarizing the system while only source qubits are engaged ($s\geq 0$); excitation dynamics are initialized when the source lines are quenched from $s\gg 1$ to $s\ll 1$.  Measurement is achieved by quenching the detector lines from $s\ll 0$ to $s\gg 1$.  Per-line delay allows fine-grained control of $t$.}
  \label{fig:ed_protocol}
\end{figure*}

Our experiments use the protocols for excitation and measurement of a static Hamiltonian laid out in Ref.~\cite{DeshpandeAnalogDigital2026}.  Protocol waveforms are shown in Fig.~\ref{fig:ed_protocol}.

\subsection{Calibration refinement} \label{subsec:calibration_refinement}

To refine the base QPU calibration, we apply similar methods to those used in Ref.~\cite{DeshpandeAnalogDigital2026,dwave_docs}.  All the methods we use operate on either three or four qubits: one target qubit with a source and a detector, or a two-qubit target dimer with a source and a detector.

We begin by identifying an approximate reference time $\tilde t_0$ (see Extended Data Fig.~\ref{fig:ed_tzero}), defined empirically as the point of maximum occupation of a $\pi$ pulse in a single target qubit's $\sigma^z$ basis, averaged over many embeddings of the single-target three-qubit system.

Working in the vicinity of $\tilde t_0$, we then coarsely synchronize the source and detector annealing lines; there are two of each in the 2D case and three of each in the 1D and 3D cases.  This is done by setting per-line delays to synchronize average flux-basis ($\tfrac{\pi}{2}$ pulse, $\sigma^x$ basis) Larmor precession curves.

Next, we tune three-qubit systems that cover the qubits used in the many-body experiment in question.  We first bring them near to degeneracy in isolation, using flux offsets.  We then iterate several operations:

\begin{enumerate}
    \item Couple the qubits together two at a time and run a two-qubit target-detector protocol, sweeping a flux offset curve to bring the qubits to degeneracy on the $\sigma^x$-$\sigma^y$ plane.
    \item Fine tune anneal offsets on the target qubits to homogenize the oscillation frequencies to the nominal $\Delta$ (either $\SI{2.0}{GHz}$ or $\SI{3.0}{GHz}$)
    \item Sweep anneal offsets on the source and detector qubits to synchronize the oscillation phases of the qubits.
\end{enumerate}

\begin{figure*}\includegraphics[width=\linewidth]{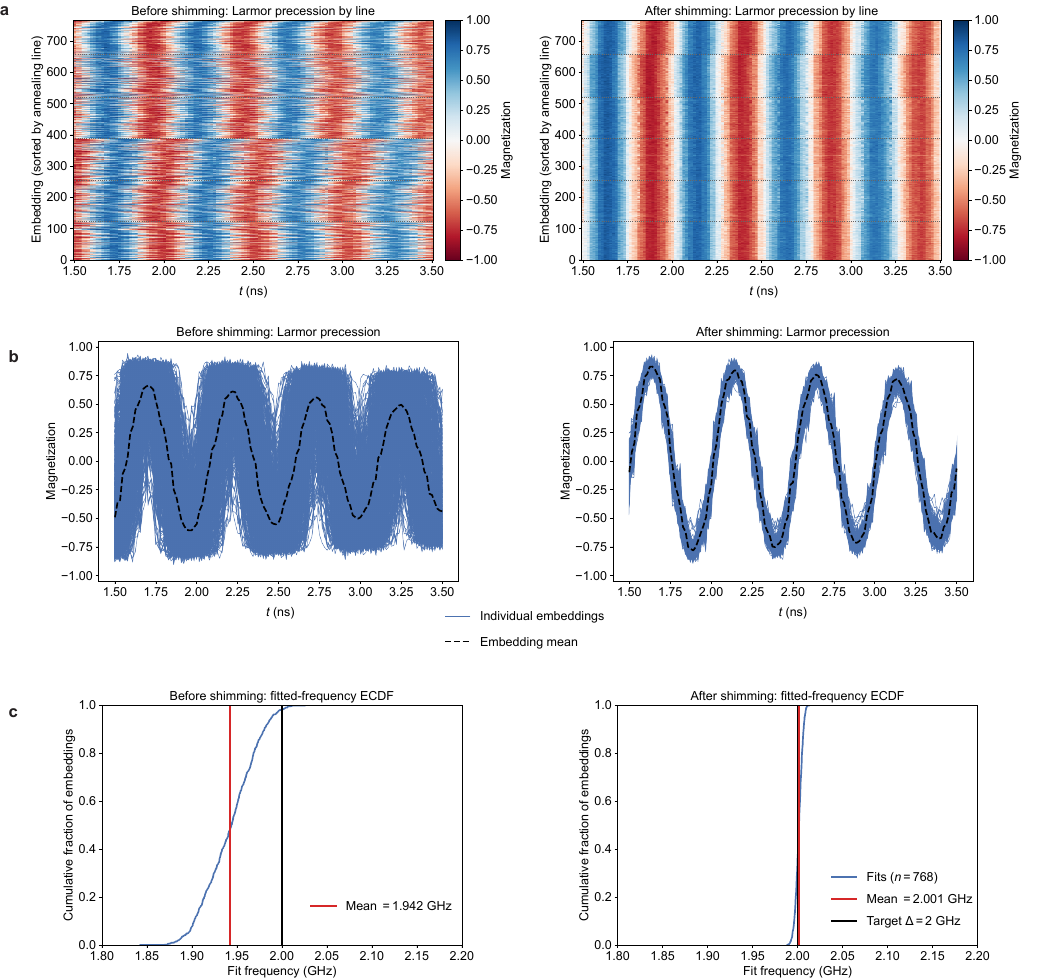}
  \caption{{\bf Calibration refinement on single-qubit systems.} Magnetization time series ({\bf a} and {\bf b}) are shown before (left) and after (right) several iterations of calibration refinement.  {\bf c}, Distribution of fitted oscillation frequencies before and after refinement.  Systems shown correspond to 12 parallel embeddings of an $L{=}32$ SSH chain, each with 64 target qubits, 64 source qubits, and 64 detector qubits.
  }
  \label{fig:sm_shim}
\end{figure*}

The distribution of Larmor precessions before and after this loop in a 1D system is shown in Fig.~\ref{fig:sm_shim}.  For dimer systems (i.e., in 2D and 3D), after single-target-qubit systems are homogenized, we couple them together in dimers and homogenize the dimer tunnelling frequencies by tuning target anneal offsets.  We do not refine the calibration any further in the many-body regime since we work near the small-coupling limit and coupler effects are therefore negligible.

\section{Extracting tight-binding model evolution from the quantum simulator}
We used an analogue evolution generated by a transverse-field Ising model (TFIM) to realize analogue evolutions for various tight-binding models (TBMs), such as the Su-Schrieffer-Heeger (SSH) model and the Benalcazar-Bernevig-Hughes (BBH) quadrupole and octupole insulator models.
This is enabled by a mapping from TFIMs to TBMs in the weak coupling regime, where the transverse field is far greater in magnitude than the Ising exchange interactions.

\subsection{Mapping a Transverse-Field Ising Model to a Tight-Binding Model} \label{eq:mapping_TFIM_TBM}
The QPU realizes a TFIM of the form
\begin{align}
    H_\text{TFIM} = -\frac{\Delta}{2} \sum_{i=1}^N  \sigma_i^z + \frac{1}{2} \sum_{i<j} J_{ij} \sigma_i^x \sigma_j^x .  \label{eq:TFIM_supp}
\end{align}
(Henceforth, we refer only to the target qubits, and exclude the auxiliary source and detector qubits from our discussion.
Additionally, we will set the Planck constant, $h$, to $1$.)
In our experiments, we choose $\Delta > 0$ and set the couplings to be much smaller in magnitude, so that $\abs{J_{ij}} \ll \Delta$ for all $(i,j)$ pairs.
In all of our experiments (except the one corresponding to Fig.~4c and Extended Fig.~5c), we assume, for simplicity, that $\Delta$ is position independent.
(The effect of a detuning, to first order, is the addition of an on-site potential.)

In this weak coupling regime, the exchange interaction terms, $V = \frac{1}{2}\sum_{i<j} J_{ij} \sigma_i^x \sigma_j^x$, can be treated perturbatively, with $H_0 = -\frac{1}{2} \Delta_i \sigma_i^z$ being the dominant term in the Hamiltonian.
The ground state of $H_0$ is the fully polarized state $\ket \Omega \equiv \ket{0}^{\otimes N}$.
The states $\ket{n} \equiv \sigma_n^x \ket \Omega$ with $n=1,\dotsc, N$ have exactly one spin flip each and span the first excited states manifold, which is well separated in energy from $\ket \Omega$.

The effective Hamiltonian within any eigensubspace (or a union of eigensubspaces) of $H_0$ can be obtained using degenerate perturbation theory~\cite{sakurai2020modern} (or the Schrieffer-Wolff transform~\cite{BravyiSchrieffer2011, coleman2015introduction}, if degeneracy is broken by site-dependent transverse-field detunings, as required for Fig.~4c) by expanding in $\norm{V}/\norm{H_0}$. 
To first order, the effective Hamiltonian is given by
\begin{align}
    H_\text{eff} = P H_0 P + P V P, \label{eq:Schrieffer_Wolff_first_order}
\end{align}
where $P$ denotes the projector onto the subspace of interest.

If the initial state is primarily in the union of the single-spin-flip and zero-spin-flip sectors (with the latter being spanned simply by $\ket{\Omega}$), the effective Hamiltonian (to first order) can be obtained by using Eq.~\eqref{eq:Schrieffer_Wolff_first_order}:
\bseq \label{eq:first_order_zero_and_one_spin_flip_sectors}
\begin{align}
    H_\text{eff}^{(01)} &= 0 \ket{\Omega}\bra{\Omega} + H_\text{TBM}, \\
    \text{where } H_\text{TBM} &\equiv  \Delta \mathbb I_1 + \tfrac{1}{2} \sum_{i<j} J_{ij} \left( \ket{i}\bra j + \ket j \bra i\right). \label{eq:H_TBM} 
\end{align}
\eseq
Here, $\mathbb I_1 = \sum_n \ket{n} \bra{n}$ is the identity operator within (or, equivalently, the projector onto) the single-spin-flip sector.
Here, we shifted the energy by $-N\Delta/2$, which has no consequence on the dynamics.
Even though the energy of the fully polarized state term is $0$, we retain it here to make the relative energy difference $\Delta$ with the single-spin-flip sector explicit.
(This relative energy difference manifests in the propagator and results in a shift of the spectrum by $\Delta$.)

If the initial state has the most weight in the single-spin-flip sector only, the relevant effective Hamiltonian is instead
\begin{align}
    H_\text{eff}^{(1)} &= H_\text{TBM},
\end{align}
where $H_\text{TBM}$ is the one defined in Eq.~\eqref{eq:H_TBM}.

Second-order virtual processes induce longer range hoppings and an additional energy offset. 
However, for the parameter ranges we consider in our work, we find that the second-order terms can be ignored, and the first-order terms dominate and suffice.
To illustrate this, let us consider a uniform chain of length $L$ with $J_{i,i+1}=J$ for all $i$.
It is straightforward to show that the second-order effective Hamiltonian (for the single-spin-flip sector) is given by
\begin{align}
    H_\text{eff} = \Delta \left( 1 + \frac{J^2}{4\Delta^2} \right) + \frac{1}{2}\sum_i J \left(\ket{i+1} \bra{i} + \ket i \bra{i+1}\right) -\frac{J^2}{8\Delta} \sum_i  (\ket{i+2} \bra i + \ket i \bra{i+2}) + \bigO \left( \frac{J^3}{\Delta^2} \right). \label{eq:second_order_SW_for_chain}
\end{align}
There are two second-order terms: a small energy offset, and next-nearest-neighbour hopping terms.
However, both are small in the parameter regimes we consider.
For example, with $J/\Delta=0.1$, the second-order offset term is $0.25\%$ of the first order contribution, while the magnitude of the next-nearest-neighbour hopping terms is about $2.5\%$ of that of the nearest-neighbour terms.

\subsection{Inferring TBM dynamics using TFIM dynamics} \label{sec:dictionary}

The extraction of topological invariants for the various models we study ultimately relies on accessing the complex amplitude $\bra{n}\ket{\psi(t)}$ as a function of time $t$ and location $n$ for a tight-binding model of the form Eq.~\eqref{eq:H_TBM} with the wavefunction initialized fully at some orbital at time $t=0$.
Since the quantum annealer natively realizes a TFIM instead of a TBM, we need a dictionary that maps measurements in the former to wavefunction amplitudes in the latter.
Thus, we develop this dictionary in this section.

For a wavefunction (in the tight-binding picture) initialized at $\ket{\psi(t=0)} = \ket{m}$, where $m$ subsumes position and the orbital index, we can access the full complex amplitude (to first order in perturbation theory) using the minimal dictionary shown in Table~\ref{tab:dict}:
\begin{table}[h!]
    \centering 
    \begin{tabular}{|c|c|}
        \hline
        Tight-Binding Model (TBM) & Transverse-Field Ising Model (TFIM) \\
        \hline
        $\bra{n}\ket{\psi(t)}$  & $2 \langle \sigma_n^-(t) \rangle_{\psi_m}$ \\
        $\Re \bra{n}\ket{\psi(t)}$ & $\expval{\sigma_n^x(t)}_{\psi_m}$ \\
        $\Im \bra{n}\ket{\psi(t)}$ & $-\expval{\sigma_n^y(t)}_{\psi_m}$ \\
        $\abs{\bra{n}\ket{\psi(t)}}^2$ & $\tfrac{1}{2} (1-\langle \sigma_n^z(t)\rangle_{\widetilde \psi_m})$\\
        \hline
    \end{tabular}
    \caption{Mapping TBM wavefunction properties to TFIM measurements.}
    \label{tab:dict}
\end{table}
Here, $\ket{\psi_m} = \frac{1}{\sqrt 2} (1+\sigma_m^x) \ket \Omega = R^y_m(\tfrac{\pi}{2}) \ket \Omega$ and $\ket{\widetilde \psi_m} = \sigma_m^x \ket \Omega$ denote states with respect to which expectation values are taken.
The second and third rows are simply the real and imaginary parts, respectively, of the first row.
While the probability $p(n,t)$ at location $n$ and time $t$ can be obtained by simply squaring and adding the real and imaginary parts of the wavefunction, said probability can also be obtained using a $\sigma^z$ measurement (fourth row of the table).
This corresponds to a quantum walk evolving under the tight-binding Hamiltonian.

We now derive the mapping in the second and third rows.
In the weak coupling limit, we may restrict the TFIM dynamics to the zero- and one-spin flip sectors (since the initial state is in the union of these sectors), so that the  wavefunction at time $t$ is given by
\bseq
\begin{align}
    \ket{\psi(t)} &= e^{-i 2\pi H_\text{TFIM}t} \ket{\psi_m} \\
    &\approx \frac{1}{\sqrt 2} (\ket \Omega + e^{-i 2\pi H_\text{TBM} t} \ket{m}).
\end{align}
\eseq
Here, we used the first-order effective Hamiltonian from Eq.~\eqref{eq:first_order_zero_and_one_spin_flip_sectors}.
It then follows that 
\begin{align}  
    \begin{split}
    \expval{\sigma_n^x(t)} &= \bra{\psi(t)} \sigma_n^x \ket{\psi(t)}
    \approx \tfrac{1}{2} \left[\bra{n} e^{-i 2\pi H_\text{TBM}t} \ket{m} + \bra{m} e^{i 2\pi H_\text{TBM} t} \ket{n}\right]\\
    &= \Re \bra{n} e^{-i2\pi H_\text{TBM}t} \ket{m} \equiv \Re \bra{n}\ket{\psi(t)}_\text{TBM},
    \end{split}
\end{align}
while 
\begin{align}
    \begin{split}
    \expval{\sigma_n^y(t)} &= \bra{\psi(t)} \sigma_n^y \ket{\psi(t)} 
    \approx \tfrac{1}{2} \left[-i\bra{n} e^{-i2\pi H_\text{TBM}t} \ket{m} +i \bra{m} e^{i 2\pi H_\text{TBM} t} \ket{n}\right] \\
    &= \Im \bra{n} e^{-i2\pi H_\text{TBM}t} \ket{m} \equiv \Im \bra{n}\ket{\psi(t)}_\text{TBM}.
    \end{split}
\end{align}

Thus, the real and imaginary parts of the wavefunction amplitude correspond to $\sigma^x$ and $\sigma^y$ measurements, respectively.

The last row of Table~\ref{tab:dict}, which pertains to the probability of the wavefunction, can be derived similarly within first-order perturbation theory applied to the single-spin-flip sector.

\section{Spectral Projector Tomography and Local Density of States}
\subsection{Band Projectors as Response Function Residues}
The Bloch projection operators can be obtained by analysing the complex wavefunction amplitude as a function of time and location, as discussed in the following section. 
However, it is useful to also view the experimental measurements in terms of Green's functions, and to then incorporate both the zero-temperature approximation and the weak-coupling limit.

The protocol for accessing the wavefunction amplitude, summarized in the first row of Table~\ref{tab:dict}, starts with the rotation $R^y(\tfrac{\pi}{2})$, is followed by an analogue evolution, and ends with a $2\sigma^+$ measurement.
($2\sigma^+$ itself is not Hermitian, so it cannot be measured directly. 
Instead, we here refer to two separate experiments: one in which we measure $\sigma^x$, and another in which we measure $\sigma^y$.
We then combine them using $2\sigma^+ = \sigma^x + i \sigma^y$.)

Although the initial state is $\ket \Omega = \ket 0 ^{\otimes N}$ to a good approximation, for now, we assume it is a thermal density matrix $\rho \propto e^{-\beta H}$. 
Later, we will take the limits $T\rightarrow 0$ and $J/\Delta \rightarrow 0$ limits, which will reestablish $\ket \Omega$ as the initial state.

Assuming that the qubits are arranged in a lattice with $N_\text{subl}$ number of sublattices, 
then, for a general initial state $\rho$, if $(i,\alpha)$ and $(j,\beta)$ denote the qubit locations (with $i,j$ denoting lattice vectors, and $\alpha,\beta \in \{1,\dotsc,N_\text{subl}\}$, the sublattice indices) of the initial rotation and the final measurement, this protocol measures
\begin{align}
    M_{\beta \alpha}^{(\rho)}(j,i,t) = 2\theta(t)\Tr[\rho R_{i,\alpha}^\dagger \sigma_{j,\beta}^+ (t) R_{i,\alpha}]. \label{eq:M_ji_plus_step_1}
\end{align}
Note that we used $\theta(t)$ because our measurements correspond to positive time values only.

This expression can be simplified.
First, we note that for any operator $O$, $R_{i,\alpha}^\dagger O R_{i,\alpha} = \tfrac{1}{2}(O + \sigma_{i,\alpha}^y I \sigma_{i,\alpha}^y + i [\sigma_{i,\alpha}^y,O])$;
Next, we note that the parity operator $\mathcal P = \prod_{i,\alpha} \sigma_{i,\alpha}^z$ commutes with $H$: $[H,\mathcal P]=0$;
$[\mathcal P,\rho]=0$;
and $\sigma_{j,\beta}^+(t)$ and $\sigma_{i,\alpha}^y \sigma_{j,\beta}^+(t) \sigma_{i,\alpha}^y$ are odd under parity.
Thus, Eq.~\eqref{eq:M_ji_plus_step_1} can be simplified to
\begin{align}
     M_{\beta\alpha}^{(\rho)}(j,i,t) = -i \theta(t) \expval{[\sigma_{j,\beta}^+(t),\sigma_{i,\alpha}^y]}_\rho.
\end{align}
This expression equals a retarded response function $M_{\beta\alpha}^{(\rho)}(j,i,t) = \chi_{\sigma_{j\beta}^+, \sigma_{i\alpha}^y}^R(t,\rho)$, where $\chi^R$ is defined as $\chi_{AB}^{R}(t;\rho) = -i \theta(t) \expval{[A(t),B(0)]}_\rho$.
Due to translational invariance, we may write
\begin{align}
    M_{\beta\alpha}^{(\rho)}(j,i,t) \equiv M_{\beta\alpha}^{(\rho)}(j-i,t).
\end{align}
Although everything is exact so far, to make analytical progress, we must introduce the two assumptions mentioned above.
First, in the weak coupling limit, the ground state is approximately $\ket \Omega$, while the first set of excited states is spanned by linear combinations of elements of $\{\ket{n,\chi}|n=1,\dotsc,N;\ \ \chi=1,\dotsc N_\text{subl}\}$.
The effective Hamiltonian in the union of the corresponding subspaces is 
\begin{align}
    H_\text{eff}^{(01)} = 0 \ket{\Omega } \bra {\Omega} + H_\text{TBM},\label{eq:HTBM_Bloch_0} 
\end{align}
as derived above (see Eq.~\eqref{eq:first_order_zero_and_one_spin_flip_sectors}).
For a translationally invariant $H_\text{TBM}$, we may use a Bloch decomposition.
Although what follows can be easily extended to higher dimensions, for simplicity, let us consider a one-dimensional system.
We may write $H_\text{TBM} = \sum_{\nu=1}^{N_\text{subl}}\sum_k E_\nu(k) \ket{k,u_\nu(k)} \bra{k,u_\nu(k)}$, where $\ket{k,u_\nu(k)} = \ket k \otimes \ket {u_\nu(k)}$ is a Bloch state corresponding to band $\nu$, with $\ket{u_\nu(k)}$ being the periodic part of the Bloch state.
We use the convention $\ket k = \frac{1}{\sqrt{L}}\sum_n e^{-ikn} \ket{n}$.
Equivalently, we may write Eq.~\eqref{eq:HTBM_Bloch_0} as
\begin{align}
    H_\text{TBM} = \sum_{\nu=1}^{N_\text{subl}}\sum_k E_\nu(k) \ket{k} \bra k \otimes P_\nu(k), \label{eq:HTBM_Bloch} 
\end{align}
where $P_\nu(k) \equiv \ket{u_\nu(k)}\bra{u_\nu(k)}$ is the projector corresponding to band $n$.
Now, let us also assume that the temperature $T=\beta^{-1}$ of the QPU is small, so that almost all the weight of the thermal density matrix lies in these two sectors, with the weight in the sectors with two or more spin flips being negligible.
That is, we assume that $\rho \approx \frac{1}{\mathcal Z} \left( \ket \Omega \bra \Omega + \sum_{k,\nu} e^{-iE_\nu(k)} \ket{k,u_\nu(k)} \bra{k,u_\nu(k)} \right)$.
After some algebra, we obtain
\begin{align}
    M_{\beta\alpha}^{(\beta)}(n,t) &= \frac{\theta(t)}{L\mathcal Z} \sum_{\nu,k} \left(1-e^{-\tfrac{E_\nu(k)}{T}} \right) e^{-i2\pi E_\nu(k)t} e^{-ikn} P_\nu^{\beta\alpha}(k).
\end{align}
Here, $P_\nu^{\beta\alpha}(k) = \bra{\beta}P_\nu(k) \ket{\alpha}$ denotes matrix elements of the projector $P_\nu$.
Setting $T \rightarrow 0$, which corresponds to $\rho_\text{init} = \ket \Omega \bra \Omega$ in the weak-coupling limit), we get
\begin{align}
    M_{\beta\alpha}(n,t) &= \frac{\theta(t)}{L\mathcal Z} \sum_{\nu,k} e^{-i2\pi E_\nu(k)t} e^{-ikn} P_\nu^{\beta\alpha}(k).
\end{align}
Taking the two-dimensional Fourier transform\footnote{\begin{minipage}{2\linewidth}{We use the convention where the two-dimensional Fourier transform is defined via $f(m,t) \rightarrow f(k,\omega) \coloneqq \sum_m \int_{-\infty}^\infty e^{i(km+\omega t)}f(m,t) \dd t$.}\end{minipage}}, we get
\begin{align}
    M_{\beta\alpha}(k,\omega) = \frac{i}{\mathcal Z}\sum_\nu \frac{P_\nu^{\beta\alpha}(k)}{\omega - 2\pi E_\nu(k) + i0^+}. \label{eq:M_gamma_nu_Lehmann_outcome}
\end{align}
It is clear that the poles in Eq.~\eqref{eq:M_gamma_nu_Lehmann_outcome} are at $\omega = 2\pi E_\nu(k)$, while the residue equals $P_\nu^{\beta\alpha}(k)$ (up to a constant), a matrix element of the Bloch projection operator.
Eq.~\eqref{eq:M_gamma_nu_Lehmann_outcome} is essentially the Lehmann representation~\cite{bruus2004many} for the TFIM, under the weak-coupling limit and a low temperature approximation (which is finally set to $0$).

\subsection{Extracting Band Projectors from Finite-Time Grid Data} \label{sec:band_structure_from_qpu}
Eq.~\eqref{eq:M_gamma_nu_Lehmann_outcome} establishes the main idea: the peaks (poles) of the Fourier-transformed QPU response are located at Bloch energies, while the residues equal the projection operators.
While Eq.~\eqref{eq:M_gamma_nu_Lehmann_outcome} predicts a well-localized pole for each band at any $k$, because the QPU data is available over a discrete grid of time points over a finite time window, we get finite-sized peaks with finite widths instead of infinitely high and infinitesimally narrow peaks.
The projector extraction therefore requires an appropriate modification, which we present here.

Henceforth, we will work exclusively in the TBM picture.
We will refer to TBM wavefunction amplitudes, which should be understood to have been obtained using the first row of Dictionary~\ref{tab:dict}.

Let $\psi^{\beta\alpha}(m,t) = \bra{m,\beta}U(t)\ket{0,\alpha}$, where $U(t)=\exp(-iHt)$ denotes the time evolution operator. 
In our experiments, we have access to $\psi^{\beta\alpha} (m,t_j)$ for $t_j=j\delta_t$ with $j=0,1,\dotsc,N_\text{t}-1$.
Taking the Fourier transform along position, we have
\begin{align}
    \psi^{\beta\alpha}(k,t_j) = \sum_{m=0}^{L-1} e^{ikm} \psi^{\beta\alpha} (m,t_j). \label{eq:psi_k_t}
\end{align}
Using the Bloch decomposition Eq.~\eqref{eq:HTBM_Bloch} and expanding $U(t)$ in terms of Bloch eigenstates, we get
\begin{align}
     \psi^{\beta \alpha}(m,t) &=  \frac{1}{L}\sum_\nu \sum_k e^{-i(km + 2\pi E_\nu(k)t)} P_\nu^{\beta \alpha}(k) \label{eq:psi_betaalpha_vanilla_in_terms_of_P}
\end{align}
Plugging into Eq.~\eqref{eq:psi_k_t}, we get
\begin{align}
    \psi^{\beta\alpha}(k,t_j) = \sum_\nu e^{-i2\pi E_\nu(k)t} P_\nu^{\beta \alpha} (k).
\end{align}
Now, we define
\begin{align}
    B_\nu^{\alpha \beta} (k) = \frac{1}{N_\text{t}} \sum_{j=0}^{N_\text{t}-1} e^{i2\pi E_\nu (k) t_j} \psi^{\alpha \beta} (k,t_j), \label{eq:def_b_nu}
\end{align}
which simplifies to 
\bseq
\begin{align}
    B_\nu^{\alpha\beta}(k) &= \sum_\chi C_{\nu\chi}(k) P_\chi^{\alpha \beta} (k) \\
    \text{where }C_{\nu\chi}(k) &= \frac{1}{N_t} \sum_{j=0}^{N_\text{t}-1} e^{i2\pi(E_\nu(k)-E_\chi(k))t_j}. \label{eq:def_C_mat}
\end{align} \label{eq:def_B_C_mats}
\eseq
$C(k)$ can be thought of as a spectral leakage matrix.
For any $\alpha,\beta$ combination, $B^{\alpha\beta}(k)$ and $P^{\alpha\beta}(k)$ are $N_\text{orb}\times 1$ vectors, while $C(k)$ is an $\Norb\times \Norb$ matrix.
Thus, we may write
\bseq
\begin{align}
    B^{\alpha\beta} &= C P^{\alpha\beta} \\
    \text{or }P^{\alpha\beta}_\nu(k) &= \sum_\chi (C^{-1}(k))_{\nu\chi} B_\chi^{\alpha\beta}(k). \label{eq:P_in_terms_of_C_B}
\end{align}
\eseq
Importantly, this expression is exact for a closed quantum evolution.

Thus, the recipe is to first extract all Bloch bands $E_\nu(k)$ using the ridges of $M(k,\omega)$ (or rather its finite-time grid analogue (see section~\ref{sec:band_structure_data_analysis} below)).
Next, we construct $B_\chi$ from Eq.~\eqref{eq:def_b_nu}, $C_{\nu\chi}$ from Eq.~\eqref{eq:def_C_mat}, and then combine them using Eq.~\eqref{eq:P_in_terms_of_C_B} to obtain $P_{\nu}$.

\subsection{Local Density of States and Fractionalization} \label{sec:LDOS_and_Frac}
In the main text, we extracted the local density of states (LDOS) for the SSH model and the BBH quadrupole and octupole moment models.
By appropriately comparing the LDOS for topological and trivial models, we inferred fractionalization at the boundaries.

Here, we will discuss the procedure for extracting the LDOS.
We consider a tight-binding (non-interacting) Hamiltonian with basis $\ket{r,\alpha}$, where $r$ denotes cell position, and $\alpha$ denotes the sublattice index.
For now, we suppress the sublattice index for simplicity, and will reinstate it later when needed.
Let $\{ E_i\}$ be the set of eigenvalues of the Hamiltonian, and $\ket{E_i}$ the corresponding eigenvector. 
(For simplicity, we assume that each energy level is non-degenerate; whereas the other case is straightforward.)

The LDOS at location $r$ is given by $\rho_r(E) = \sum_{E_i} \abs{\bra{r}\ket{E_i}}^2 \delta(E-E_i)$.
The LDOS satisfies a sum rule: at each location $r$,
\begin{align}
    1 = \int_{-\infty}^{\infty} \dd E\ \rho_r(E). \label{eq:local_sum_rule}
\end{align}

The LDOS $\rho_r(E)$ can be extracted from the return amplitude (or the local autocorrelation function), which is defined as
\begin{align}
    A_r(t) = \bra{r} e^{-i2\pi Ht} \ket{r}. \label{eq:return_amplitude}
\end{align}
It is straightforward to show that $A_r(t)$ and $\rho_r(E)$ are related to each other via a time Fourier transform:
\begin{align}
    A_r(t) = \int \dd E\ \rho_r(E) e^{-i2\pi Et}; \qquad 
    \rho_r(E) = \int_{-\infty}^\infty \dd t \ e^{i2\pi Et} A_r(t). \label{eq:rho_r_is_FT_of_A}
\end{align}
However, since we have access only to times $t>0$, we must use the (manifestly real) form,
\begin{align}
   \rho_r(E) = \frac{1}{2} \Re \left[ \int_0^{\infty} e^{i2\pi Et} A_r(t) \dd t \right].\label{eq:rho_r_is_FT_of_A_re}  
\end{align}

Naturally, the LDOS $\rho_r(E)$ is non-zero only at the values of $E$ which are Hamiltonian eigenstates.
The LDOS can be used to infer charge fractionalization.

In the context of topological insulators, fractionalization refers to the distribution of an excess change in one many-electron system when compared against the distribution in another system~\cite{Mudry_Chamon_2025}.
(Alternatively, one may consider a ``filling anomaly'' for crystalline topological insulators~\cite{BenalcazarQuantization2019}, which can arise due to a mismatch between the number of electrons required for charge neutrality and the preservation of crystalline symmetries.)
For non-interacting topological insulators, one may compare the filled Fermi sea for a topological open system against either that for a topological periodic system, or a trivial open one (but with the same chemical potential $\mu$).
Due to constraints imposed by qubit connectivity in the QPU (in particular, the lack of fully periodic embeddings for two- and three-dimensional lattices), we compare a topological ($\gamma{<}\lambda$) open system against a trivial ($\gamma{>}\lambda$) open system, and $\gamma$ (intracellular hopping) and $\lambda$ (intercell hopping) swapped for the two cases.

\section{Data Analysis}
As discussed above, we use a dictionary to encode wavefunction propagation in TBMs to corresponding TFIM experiments implemented in the QPU.
The TBM wavefunction propagation data can then be used to extract various properties such as the local density of states, band structure, band projectors, and, ultimately, the topological invariants.
While we employ calibration refinement (see Section~\ref{subsec:calibration_refinement}) to reduce device non-idealities, we also use appropriate data analysis techniques to map measurements from the QPU to TBM dynamics.
We discuss these techniques in this section.

As mentioned in the Methods section, the data analysis starts with identification of a $t=0$ point, which we discuss in more detail in section~\ref{sec:extracting_t=0}.
The extraction of the LDOS, the band structure (section~\ref{sec:band_structure_data_analysis}) and the projectors (section~\ref{sec:projector_extraction_data_analysis}) are then discussed in that sequence.

\subsection{Defining \texorpdfstring{$t=0$}{t=0} Using LDOS Sum-Rule Diagnostic} \label{sec:extracting_t=0}

All the quantities we extract rely on accessing the wavefunction amplitude at different times in a TBM, with the wavefunction initialized at one orbital.
For example, we require the return amplitude $A_r(t)$ (as defined in Eq.~\eqref{eq:return_amplitude}) to extract the LDOS $\rho_r(E)$ (see Eq.~\eqref{eq:rho_r_is_FT_of_A_re}), and the wavefunction propagation $\psi^{\beta,\alpha}(m,t) = \bra{m,\beta}U(t)\ket{0,\alpha}$ to obtain the band projectors via Eq.~\eqref{eq:P_in_terms_of_C_B}.  Here, $t$ denotes the length of the analogue evolution in Fig.~\ref{fig:1}b (cf.~Ref.~\cite{DeshpandeAnalogDigital2026} Fig.~1d).  In the QPU's programming interface, the corresponding input parameter is $t_\text{prog}$, which controls a delay on the detector qubits' annealing-line waveforms relative to the source and target qubits~(Fig.~\ref{fig:ed_protocol}).  However, the relationship between $t_\text{prog}$ and $t$ is subject to a significant processor-dependent offset; here, we detail the definition of the processor-independent reference point $t_0$ for which $t=t_\text{delay}-t_0$.

We first align our measurements relative to the coarsely estimated reference time $\tilde t_0$ described in Section~\ref{subsec:calibration_refinement}.  For zero-dimensional reference measurements in Figures~\ref{fig:ed_protocol},~\ref{fig:sm_shim} and \ref{fig:ed_dimer}, we set $t_0=\tilde t_0$, for simplicity.  Correctly identifying $t_0$ is crucial for phase-dependent measurements such as LDOS.  We therefore use the following technique to determine $t_0$.  The first step is to implement a phase alignment so that $\psi(t_0)$ is real---in fact, we require that $\ket{\psi(t_0)} = \ket m$ for some orbital $m$ in the TBM language.

For appropriately chosen $t_0$, we expect that exciting qubit $r$ yields a return amplitude $A_r(t_0)$ with a magnitude near unity, i.e., $|\langle 2\sigma_r^+(t_0)\rangle|\approx 1$.  We therefore compute the return amplitude using a sliding window of fixed duration $t_\text{extent}$ (e.g., in the 1D case, $\SI{20}{ns}$) starting at every available time between $\tilde t_0-\SI{1}{ns}$ and $\tilde t_0+\SI{1}{ns}$.  

For each choice of $t_0$ and for each qubit $r$,
\begin{enumerate}
    \item Construct the sequence $\{\tilde A_r(t) \mid t_0 \leq t \leq t_0+t_\text{extent}\}$, where $\tilde A_r(t) = e^{-i\phi}\langle 2\sigma_r^+(t)\rangle$ and $\phi$ is the phase of $A_r(t_0)$.
    \item Compute the inverse discrete Fourier transform of $\tilde A_r(t)$ over the time window $\{t_0 \leq t \leq t_0+t_\text{extent}\}$ to obtain $\tilde \rho_r(E)=\int \dd t e^{i2\pi Et} \tilde A_r(t)$.
    \item Integrate $\tilde \rho_r(E)$ to obtain the ``sum-rule diagnostic'' $\tilde S_r = \int \tilde \rho_r(E) \dd E$.
\end{enumerate}
We choose the $t_0$ that maximizes the qubit-averaged $\tilde S_r$ in the trivial phase ($\gamma{>}\lambda$), maximizing adherence to the sum rule  Eq.~\eqref{eq:local_sum_rule}.  While the integral should in principle be taken over all frequencies, for practical purposes, we restrict the integral to a region around $\Delta$, corresponding to the limits of the plotted LDOS.

\begin{figure*}
  \includegraphics[width=\textwidth]{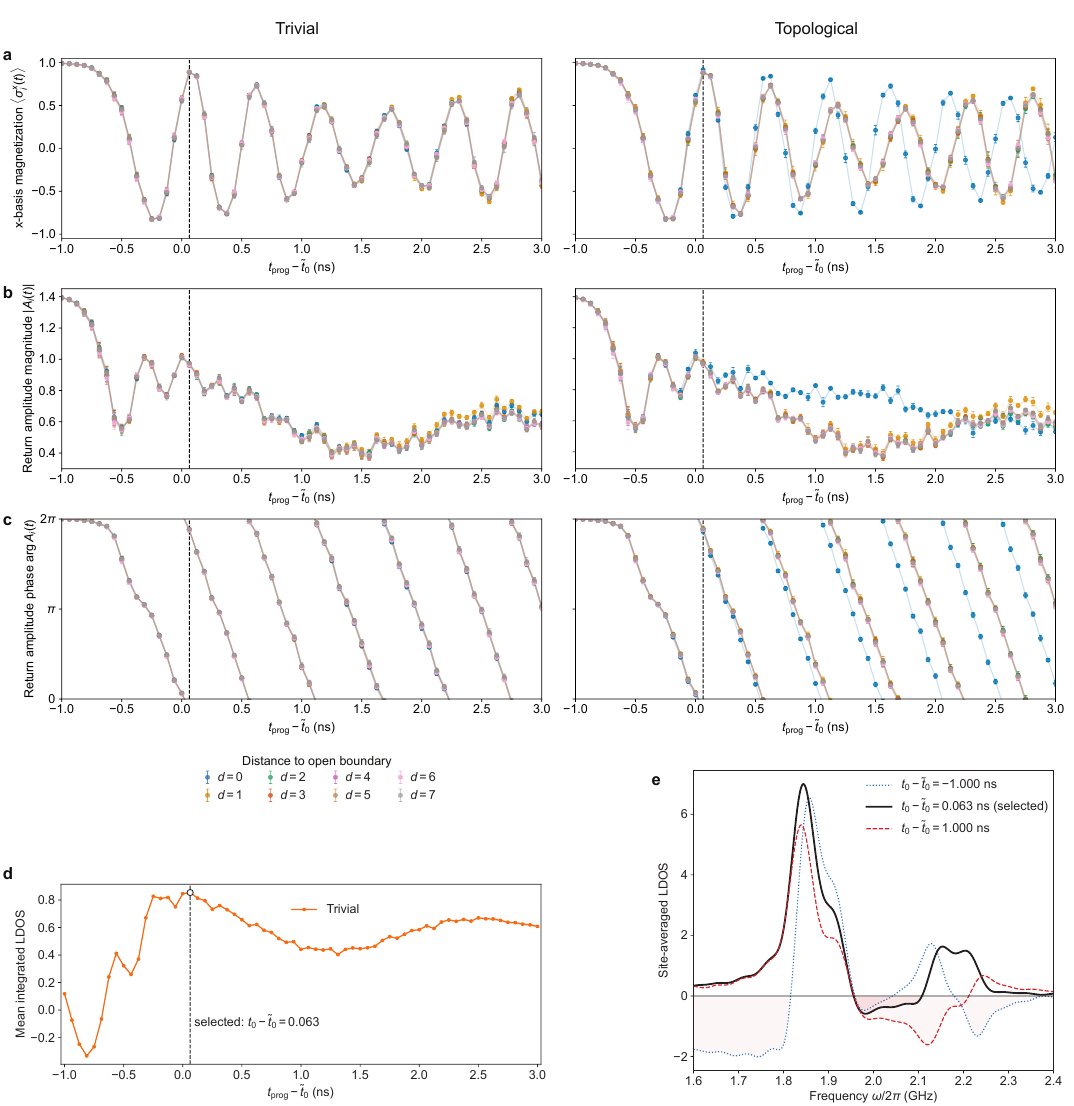}
  \caption{{\bf Determining $t{=}0$.}  Fine-grained analysis of data represented in Fig.~\ref{fig:2}d ($L{=}8$). We refine a coarse $t{=}0$ estimate from $\tilde t_0$ to the refined estimate $t_0$; $\tilde t_0$ is extracted from the maximum occupation following a $\pi$ pulse on isolated qubits.
  {\bf a}, Direct QPU measurements yield $\sigma^x$ magnetizations ($L{=}16$).  Measurements are grouped and bootstrapped according to the distance from the nearest open boundary; error bars represent 95\% confidence intervals on the mean.  Boundary sites deviate significantly from the bulk only in the topological phase.  {\bf b--c}, By time-shifting according to Eq.~\eqref{eq:timeshift} we can reconstruct complex return amplitude, whose magnitude and phase shown in {\bf b} and {\bf c} respectively.  {\bf d}, We start the analysis window at each time step in the vicinity of $\tilde t_0$ and compute the site-averaged integrated LDOS ({\bf e}).  We select the $t_0$ that maximizes the integrated LDOS in the trivial phase. {\bf e}, choosing $t_0$ too early or too late results in significant negative LDOS.  We truncate negative LDOS instead of removing it with data windowing so as to retain as much frequency resolution as possible.  Selection of $t_0$ is done once for each dimensionality, using BBH models in 2D and 3D.}
  \label{fig:ed_tzero}
\end{figure*}

The relevant measurements are shown in Fig.~\ref{fig:ed_tzero}, where we show $t_0$ candidates ranging from $\tilde t_0-\SI{1}{ns}$ t0 $\tilde t_0+\SI{3}{ns}$ in Fig.~\ref{fig:ed_tzero}d for illustrative purposes.

We compute $t_0$ for each dimensionality, finding $t_0 = \tilde t_0+\tfrac{1}{16}\SI{}{ns}$ for 1D (where we do not use dimerization) and $t_0 = \tilde t_0-\tfrac{1}{4}\SI{}{ns}$ for 2D and 3D (for both of which, we use dimerization).  When computing phase-dependent quantities such as LDOS, we perform the aforementioned phase alignment.

\subsection{Fractionalization and LDOS}

To probe fractionalization, we compute the LDOS as in the previous section. 
In particular, we use the normalized sum rule diagnostic to get the LDOS: $\rho_r(E) \leftarrow \widetilde \rho_r(E)/\widetilde S_r$.
Next, we obtain the central frequency $\tilde \Delta$ as described in Methods.  Since each model has chiral/sublattice symmetry, i.e., $\{H_\text{eff}^{(1)} - \Delta, \Gamma \} = 0$ for an appropriate chiral symmetry operator $\Gamma$ satisfying $\Gamma^2=\mathcal I$, the spectrum as well as the LDOS are expected to be symmetric about $\tilde \Delta$.
In our experiments, we find that the intensity of the LDOS decreases with increasing frequency---this is qualitatively explained by open quantum system effects, but has also been observed in acoustic metamaterials~\cite{serra-garcia_observation_2018}.
To mitigate this, we symmetrize the LDOS about the extracted value $\tilde \Delta$ when inferring fractionalization through charge deficit; LDOS plots retain the unsymmetrized values.
The chemical potential $\mu$ is set to a frequency value below the lowermost boundary mode, but above the bulk energies.
The charge deficit (in the corresponding topological phase when compared against the trivial phase) at each location $r$ is 
\begin{align}
    \rho^\text{deficit}_r = \int_{-\infty}^\infty  \dd E \ \rho_r^\text{trivial}(E) -  \int_{-\infty}^\infty  \dd E \ \rho_r^\text{topological}(E),
\end{align}
where the superscripts `topological' and `trivial' denote phases realized with open boundaries and with the same but swapped values of $\gamma$ (intracellular hopping) and $\lambda$ (intercellular hopping).
(As before, we integrate over roughly twice the frequency range that covers the bands.)

\subsection{Grid of Quasi-Momentum Values} \label{subsec:grid_quasi_momentum}
For extracting the band structure and the band projectors as well as, ultimately, the topological invariants, we extract $k$-space quantities.
Due to the finite system sizes of the models we realize on the device, we have access to only a finite grid of quasi-momentum ($\vec k$) values.

For the corresponding experiments for the SSH model in one dimension, embed chains of length $L=32$ cells (i.e., with $64$ qubits) with periodic boundary conditions.
All the information is then extracted over the grid $k_i = -\pi + \frac{2\pi i}{L}$, with $i=0,\dotsc,L$.

For the two-dimensional quadrupole insulator model (the BBH model), we use data from a skew-square lattice shown in Extended Data Figure~2, with $364$ logical qubits, each of which corresponds to an orbital in the BBH model.
The realized grid is therefore 13 unit cells wide along the diagonal direction.
(Due to the constraints of the connectivity graph of the QPU, we have open boundaries along both directions.) 
To mimic experiments corresponding to periodic boundaries, we excite the orbitals only at the central unit cell, and restrict data analysis to a shorter time window so that the excitation does not reach the boundary.
To sample at $\vec k$ points spaced in a rectangular grid in the Brillouin zone, we port the obtained real-space data to a rectangular grid with $13\times 13$ cells, setting the values at sites that are not present in the experiment to $0$.
Taking the fast Fourier transforms (FFTs) of the appropriate data over this grid along the two space directions yields a momentum grid $(k_x,k_y)$ with $k_x$ and $k_y$ assuming values $2\pi n/13$ with $n=-6,-5,\dotsc,6$.

For the three-dimensional models, we use a cubic lattice with $3\times 3\times 3$ unit cells, corresponding to a $k$-space grid with $3\times 3\times 3$ points, with equal spacing along each direction, and with the origin as one of the points.

\subsection{Band Structure Extraction} \label{sec:band_structure_data_analysis}
The poles of $M_{\gamma,\nu}(k,\omega)$ (see Eq.~\eqref{eq:M_gamma_nu_Lehmann_outcome}) coincide with the Bloch bands $2\pi E_\mu(\vec k)$ under the approximations of weak-coupling limit and closed quantum evolution.
However, Eq.~\eqref{eq:M_gamma_nu_Lehmann_outcome} implicitly assumes knowledge of wavefunction propagation over the time range $t\in[0,\infty)$.
In our experiments, we instead propagate excitations over a finite time window with a finite grid spacing.
The former restricts the range of frequencies through the Nyquist sampling theorem, and the latter results in a discretization of the frequency grid.
We define $f(\vec k,\omega) = \sum_{\nu} \abs{M_{\nu,\nu}(k,\omega)}^2$, and analyse its profile to extract $E_\mu(k)$ for the various bands $\mu$.
$M(k,\omega)$ is obtained using a two-dimensional FFT of the data $M(\vec r,t)$, which is itself obtained by averaging over multiple embeddings (in the 1D case) and available lattice symmetries (in all cases).

We now discuss the various steps implemented to extract the band structure.
First, we note that due to relaxation, dephasing, and other open quantum system effects, we consistently find that $f(\vec k,\omega)$ has an exponentially decaying profile with frequency.
To mitigate this, we exploit the fact that the tight-binding Hamiltonian is chiral-symmetric (to first order in perturbation theory)---this is evident from the lattice structure without needing detailed prior knowledge of the system.  As a result, the band structure is guaranteed to be symmetric about $\Delta$ in the weak-coupling limit; we therefore average the Fourier transform with its reflection about $\omega=2\pi\tilde\Delta$, where $\tilde\Delta$ is the empirically determined corner-mode energy.  Lattice symmetry also allows us to average the FT horizontally about $k=0$.

Second, the finiteness of the time window imposes a restriction on the frequency resolution.
To obtain a finer and smoother variation of $f(\vec k,\omega)$ with $\omega$, we implement a multi-resolution smoothing procedure.
In particular, we assume that the data from the QPU is available over an equally-spaced time grid from $t=0$ to $t_\text{extent}$.
We compute the FFT for up to 10 time windows, successively discarding the final time step.  Since the frequencies at which the FFTs are computed are different for each window, we can combine the data to get $f$ over a finer (though no longer equally spaced) frequency grid than that available from the original time window.  Although this does not add extra spectral information, it aids in the fitting described below.

We next extract $\{ E_\mu(k)\}$ using the data obtained from this multi-resolution construction.
For the 1D SSH model, the 2D quadrupole insulator (the BBH model), and the 3D octupole model, we have 2, 4, and 8 bands, respectively.
However, in all these cases, there are only two distinct Bloch energies at any $k$ value---the three cases have band degeneracies of 1, 2, and 4 respectively.
The degeneracy in the quadrupole insulator case is fully explained by the combination of inversion symmetry $\mathcal P =\mathcal M_x \mathcal M_y$---with $\mathcal M$s denoting mirror symmetries---and time-reversal symmetry $\mathcal T$: the symmetry $\mathcal P \mathcal T$ satisfies $(\mathcal P \mathcal T)^2=-1$, implying a two-fold degeneracy of each energy level.
For the octupole insulator, a two-fold degeneracy follows similarly with the inversion symmetry $\mathcal P= \mathcal M_x \mathcal M_y \mathcal M_z$.
The four-fold degeneracy is explained by expressing the Hamiltonian in terms of the $\Gamma$ matrices~\cite{BenalcazarQuantized2017} and squaring it.

Consistent with this, we observe that the function $f$ obtained from the QPU data also has only two prominent peaks at each quasi-momentum value, converging to one at a closing gap.
To extract the two band energies as a function of $k$, we implement a sum-of-two-Lorentzians fit at each $\vec k$ independently.
In particular, we fit $f_{\vec k}(\omega)$ to the form $g(\omega) = a \gamma^2  \left(\frac{1}{(\omega-\mu_1)^2 + \gamma^2} + \frac{1}{(\omega-\mu_2)^2 + \gamma^2} \right)$, with fitting parameters $a, \Omega_\text{lower}, \Omega_\text{upper}$ and $\sigma$.
$\Omega_\text{lower}$ and $\Omega_\text{upper}$ as a function of crystal momentum are the extracted Bloch energies.

\begin{figure*}
  \includegraphics[width=\textwidth]{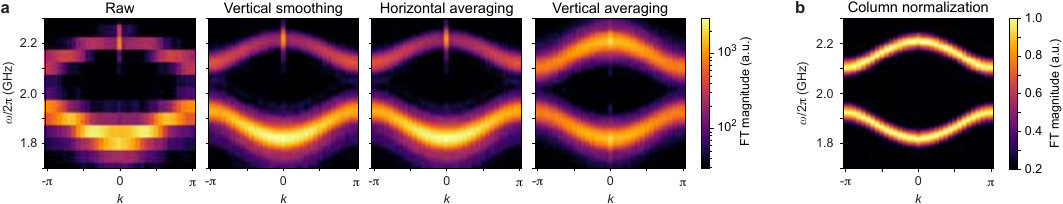}
  \caption{{\bf Processing of Fourier transform visualizations.} {\bf a}, Raw band structures are taken from fast Fourier transform over the entire time series.  Vertical smoothing is achieved by progressively shortening the time series and combining extracted data. Horizontal averaging about $k=0$ and vertical averaging about $\tilde \Delta$ are then applied.  {\bf b}, Finally, each column of the Fourier transform is normalized, giving the 1D SSH band structure shown in Fig.~\ref{fig:2}c.}
  \label{fig:ed_fouriertransforms}
\end{figure*}

In the main text, for visual clarity, we normalize the $f(\vec k,\omega)$ heatmaps so that each column (i.e., each $\vec k$ value) ranges between $0$ and $1$.  The successive treatments of the FT are shown for a 1D example in Fig.~\ref{fig:ed_fouriertransforms}.  We do not expect them to significantly alter extracted band-level fits.

\subsubsection{Adequacy of finite systems with open boundaries}
Accurate extraction of Bloch energies is required for spectral projector tomography, since the spectral-leakage matrix $C$ depends explicitly on the band energies; see Eq.~\eqref{eq:def_B_C_mats}.
Bloch bands are defined for translationally invariant lattices, while a finite system with periodic boundary conditions (PBCs) provides discrete samples of the corresponding dispersion over the allowed crystal momenta.

For the SSH model, we realize periodic chains with $L=32$ unit cells, giving $32$ uniformly spaced crystal momenta across the Brillouin zone.
This resolution is sufficient to resolve the two SSH bands and their evolution across the topological transition.

In two and three dimensions, the connectivity constraints of the QPU make fully periodic embeddings considerably more restrictive.
In particular, auxiliary source and detector qubits must be reserved and connected to the target lattice, preventing us from imposing PBCs along every spatial direction.
The two-dimensional lattices therefore have open boundaries along both directions, while the three-dimensional BBH lattice contains $3\times3\times3$ unit cells and is periodic only along $z$.

We mitigate the resulting finite-boundary effects by initializing excitations near the centre of the lattice and restricting the analysis to short evolution times.
Before appreciable amplitude propagates to the boundary and returns to the bulk region, the local propagator closely approximates that of the corresponding translationally invariant system.
The Fourier-transformed response can therefore be used to estimate the bulk dispersion even when the experimentally realized lattice is not fully periodic.

We verify this approximation directly by comparing the experimentally extracted band energies with the exact Bloch dispersion.
Since the spectral-leakage matrix depends only on differences between band energies, we quantify the accuracy of the extracted dispersion using the error in the band splitting rather than the absolute frequencies.
For the models considered here, there are two distinct energies at each crystal momentum, so we define
\begin{align}
    \Delta \epsilon(\vec k)
    =\frac{
    E_\text{upper}(\vec k)-E_\text{lower}(\vec k)}{E_\text{upper}(\vec 0)-E_\text{lower}(\vec 0)}.
    \label{eq:band_splitting}
\end{align}
The normalization term is the maximum splitting, i.e., ~$2(\gamma+\lambda)$, $2\sqrt{2}(\gamma+\lambda)$ and $2\sqrt3(\gamma+\lambda)$ in 1D, 2D, and 3D, respectively.  For each combination of $(\gamma,\lambda)$, we compute the normalized root-mean-square error across the high-symmetry path points represented in the band structure plots (Fig.~\ref{fig:2}c, Fig.~\ref{fig:3}c, and Fig.~\ref{fig:5}a):
\bseq
\begin{align}
    \epsilon_\text{max} &= \max_{\vec k} \abs{\Delta \epsilon^\text{extracted}(\vec k) - \Delta \epsilon^\text{exact}(\vec k)}, \\
    \epsilon_\text{path} &=  \sqrt{\frac{1}{N_k}\sum_{\vec k}\left[\Delta \epsilon^\text{extracted}(\vec k) - \Delta \epsilon^\text{exact}(\vec k)\right]^2}.
    \label{eq:band_energy_error}
\end{align}
\eseq

\begin{figure}
  \includegraphics[scale=1]{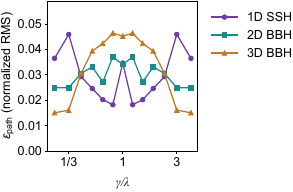}
  \caption{\textbf{QPU Extracted Band Error.} Error in band splitting relative to maximum at $\Gamma=\vec 0$ for the 1D SSH model and 2D and 3D BBH models. RMS error is taken across discrete high-symmetry path points.}
  \label{fig:sm_banderror}
\end{figure}

This dimensionless metric is insensitive to a common energy offset and directly quantifies the errors relevant to the construction of the spectral-leakage matrix. Fig.~\ref{fig:sm_banderror} shows the resulting error as a function of $\gamma/\lambda$ for the one-dimensional SSH model and the two- and three-dimensional BBH models. Despite the increasingly restrictive boundary conditions and the small linear size of the three-dimensional realization, the extracted dispersions remain close to the exact bulk bands over the parameter range used for subsequent projector reconstruction.

\subsection{Projector Extraction} \label{sec:projector_extraction_data_analysis}
Having obtained the band energies $\{E_\mu(\vec k)\}$ as described in the previous section, we plug those values into the expressions Eq.~\eqref{eq:def_B_C_mats} to obtain tensor $B$ and matrix $C$ as a function of $\vec k$.
(We also normalize the wavefunction $\psi(t)$ at each time slice when computing $\psi(\vec k,t)$ using Eq.~\eqref{eq:psi_k_t}.)
At each value of $\vec k$, we approximate the inverse of $C$ using the Moore-Penrose pesudo-inverse~\cite{strang2012linear} with a cutoff of $10^{-10}$ for the singular values of $C$.
This ensures that the projector is computed even close to those $\vec k$ points where the band gap is small, or zero, which happens at the phase transition point $\gamma{=}\lambda$, resulting in $C$ becoming singular. 
(The errors in the computation of the topological invariants also increase as we approach the transition point, reflecting the inaccuracies in computing the inverse of $C$, and consequently also the projector $P$.)

After using Eq.~\eqref{eq:P_in_terms_of_C_B} to compute $P_\nu(k)$ for each band\footnote{\begin{minipage}{2\linewidth}{When the bands are degenerate, such as in the two- and three-dimensional models, we instead obtain the projector onto each \textit{degenerate} subspace.}\end{minipage}}, we Hermitianize the obtained matrix before computing the topological invariants.

\begin{figure}
  \includegraphics[scale=1]{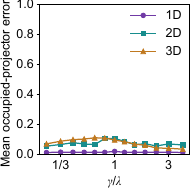}
  \caption{\textbf{QPU Extracted Projector Error.} We compute the error in the extracted lower-band projector for the SSH model and 2D and 3D BBH models, using Eq.~\eqref{eq:projector_error}. The errors (mean over Brillouin zone) are small for all values of $\gamma/\lambda$.}
  \label{fig:sm_projector_error}
\end{figure}

As discussed above, the errors in the band structure extraction are small.
To evaluate the accuracy of spectral projector tomography when applied to the QPU data, we plot the error between the filled band projector against the exact theory value for each of the models, and normalize it so that the error goes from 0 to 1.
In particular, we compute
\begin{align}
    \epsilon(\vec k) = \frac{1}{\sqrt{2d}} \norm{P_\text{extr}(\vec k) - P_\text{exact}(\vec k)}_F \label{eq:projector_error}
\end{align}
as a function of $\vec k$, where $\norm{\cdot}_F$ denotes the Frobenius matrix norm, and $d$ denotes the band degeneracy (which equals 1,2 and 4 for the occupied bands for the SSH, 2D BBH and 3D octupole BBH models, respectively) or equivalently the rank of the projector.
The factor of $\sqrt{2d}$ ensures that $\epsilon\in[0,1]$.
The Brillouin-zone-averaged error as a function of the parameters $\gamma/\lambda$ for these three models is shown in Fig.~\ref{fig:sm_projector_error}.
We consistently find that the errors are close to $0$.
In spite of the fact that the 2D and 3D model realizations do not have periodic boundary conditions along all directions, we find that the projector errors are nonetheless small.
The errors also increase as $\tfrac\gamma\lambda\rightarrow 1$, since the condition number of the leakage matrix blows up at the band gap closing points. 

\subsection{Topological Invariant Extraction}
For the SSH chain, we compute the Zak phase (which evaluates to $0$ in the trivial phase and $\pi$ in the topological phase), which is defined as
\begin{align}
    \gamma = \int_{-\pi}^\pi i \bra{u_0(k)}\ket{\partial_k u_0(k)} \dd k \mod 2\pi. \label{eq:Zak_phase_integral_1d}
\end{align}
Here, $\ket{u_0}$ denotes the periodic part of the lower band Bloch wavefunction, and is set to the eigenvector of the extracted $P_0(k)$ corresponding to the larger eigenvalue (which equals $1$ in the ideal case).
Since we use finite periodic chains in our experiments, we obtain $\ket{u(k)}$ only over a grid of $k$ values. 
Hence, we approximate Eq.~\eqref{eq:Zak_phase_integral_1d} using
\begin{align}
    \gamma = -\Im \log \prod_i \frac{\bra{u_0(k_i)}\ket{u_0(k_{i+1})}}{\abs{\bra{u_0(k_i)}\ket{u_0(k_{i+1})}}}, \label{eq:zak_phase_discrete_k}
\end{align}
with $k_i \in \left\{-\pi + \frac{2\pi x}{L}| x=1,\dotsc, L \right\}$.

For the 2D SSH model, we simply compute the Zak phase of the lowermost band (using Eq.~\eqref{eq:zak_phase_discrete_k}) along both directions, and then report the average along the transverse direction to get the polarization $p$. 
$p_y$ as a function of $\gamma/\lambda$ is reported in Extended Data Figure~5e.

For the 2D quadrupole insulator BBH model and the 3D octupole insulator model, we compute second and third nested Wilson loops corresponding to the lower bands, respectively.
Using the methods described above, we obtain the projector onto the lower two and lower four bands, which are degenerate.
We follow the procedure from Ref.~\cite{BenalcazarElectric2017} to compute the Wilson loops from our discrete momentum grids.
To compute the first Wilson loop along $x$, we first compute the overlap matrix,
\begin{align}
    M_x(k_x^{(j)}, k_\perp) \equiv V^\dagger(k_x^{(j)}, k_\perp) V(k_x^{(j+1)}, k_\perp),
\end{align}
where $k_\perp$ denotes the transverse direction, so that it equals $k_y$ in 2D and $(k_y,k_z)$ in 3D.
$V(k_x,k_\perp)$ denotes a $2\times 2$ matrix in the quadrupole case, and a $4\times 4$ matrix in the octupole case with columns being orthogonal vectors that span the occupied subspace.
These vectors are obtained from the experimentally obtained projectors.
If $M_x(k_x^{(j)}, ky) = W \Sigma U^\dagger$ denotes the singular value decomposition of $M_x$, we define its unitary part as $\widetilde M_x(k_x^{(j)}, ky) = WU^\dagger$.
The Wilson loop is computed as
\begin{align}
    W_x(k_\perp) &= \prod_{j=0}^{N_x-1} \widetilde M_x(k_x^{(j)}, k_\perp). \label{eq:W_x}
\end{align}
The Wannier centres $\nu_x(k_y)$ are obtained using the logarithm of the eigenvalues of the Wilson loop, with the convention:
\begin{align}
    W_x(k_\perp) \ket{\nu_x(k_\perp)} = e^{-i2\pi\nu_x(k_\perp)} \ket{\nu_x(k_\perp)}. 
\end{align}
The nested Wilson loop is computed by repeating the procedure for the lower Wannier bands along one of the transverse directions.

For a continuum of $k$ values and a perfect realization of the SSH model, the value is guaranteed to be exactly equal to $\pi$ or $0$. 
However, we observe small deviations from these exact values (see Fig.~2c) due to a combination of experimental errors, open quantum effects, and a weak breaking of chiral symmetry (due to second order terms in the Schrieffer-Wolff transform (see Eq.~\eqref{eq:second_order_SW_for_chain})).

In the main matter, the plotted topological invariant distributions are bootstrap populations derived from binning all symmetry- and gauge-related experiments.  For example, in 2D, the central four excitation sites are all symmetry-equivalent; thus, we derive the data for each orbital from the set of four physical excitation locations.  This approach suppresses hardware nonidealities, averaging away unwanted spatial disorder.  However, topological protection should render the material insensitive to weak disorder, so we also want to compute the invariants without using symmetries to virtually map one orbital to another.

\begin{figure*}
  \includegraphics[scale=1]{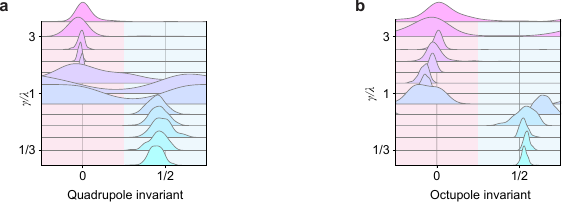}
  \caption{{\bf Invariants calculated with fewer symmetries.} We reproduce topological invariant calculations from QPU data without the $x$ and $y$ symmetries used in the main-matter figures: reflection about $x$, reflection about $y$, and exchange of $x$ and $y$. {\bf a}, 2D BBH quadrupole invariants. {\bf b}, 3D BBH octupole invariants.  Bootstrap distributions (see text) become broader and less quantized about topological values $0$ and $1/2$ but show the same qualitative phase transition.}
  \label{fig:ed_symmetries}
\end{figure*}

To this end, Fig.~\ref{fig:ed_symmetries} shows distributions of 2D and 3D BBH topological invariants corresponding to Fig.~\ref{fig:3}g and Fig.~\ref{fig:5}c, respectively.  In 2D, we excite at the 12 central sites and map, via translation and zero padding, each site to its most central site in the same orbital; this allows us to bootstrap over pools of size three without enforcing lattice symmetries such as reflection or exchange of $x$ or $y$.  In 3D, we retain only the six mappings that rotate about (by an even number of steps) or reverse the periodic $z$ dimension, giving a bootstrap pool of six QPU experiments for each orbital.

In both cases, we see that the invariants remain centred near the same topological values as in the symmetry-exploiting case.  This indicates that while orbital-to-orbital mappings quantize the invariants and tighten experimental uncertainties, they are not required to obtain the correct topological values.

\subsubsection{Stability of topological invariant extraction in 3D}
As shown above, the extracted band energies and projectors remain accurate despite the absence of periodic boundary conditions along all spatial directions in the two- and three-dimensional realizations.
However, the extraction of topological invariants introduces an additional finite-size constraint because the bulk invariants are defined over the Brillouin zone whereas the experiment samples only a discrete grid of momenta.
This discretization is relatively fine for the one- and two-dimensional systems, with $32$ and $13\times13$ momentum points, respectively, but said discretization is particularly coarse for the three-dimensional BBH model, where the $3\times3\times3$ lattice provides only three samples along each reciprocal-space direction.

To determine whether this resolution is sufficient to characterize an octupole phase, we apply the complete experimental analysis pipeline to ideal tight-binding simulations, using the same lattice geometry, boundary conditions, parameter values, and finite time window as in the QPU experiments.
Thus, the simulated data are subjected to the same band-structure extraction, spectral-projector reconstruction, and nested-Wilson-loop analysis used for the experimental measurements.

\begin{figure*}
  \includegraphics[width=\textwidth]{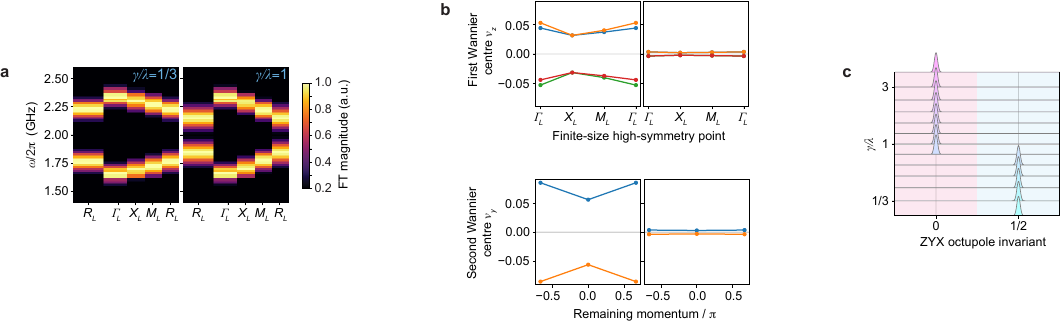}
  \caption{{\bf Tight-binding simulations of 3D octupole BBH model with extraction of topological invariant.} {\bf a}, Spectra obtained from an $L{=}3$ lattice with open $x$ and $y$ boundaries and periodic $z$ boundaries and a time-series extent of $\SI{6}{ns}$, reflecting quantum experiments.  {\bf b}, First and second Wannier centres. {\bf c}, Final octupole invariants; finite width of distributions is an artifact of smoothing.}
  \label{fig:ed_tightbinding}
\end{figure*}

The resulting finite-size Wannier spectra differ appreciably from their thermodynamic-limit counterparts, as expected from the coarse momentum grid.
Nevertheless, the final nested polarization clearly distinguishes the two phases: it remains close to $1/2$ for $\gamma{<}\lambda$ and close to zero for $\gamma{>}\lambda$, as shown in Fig.~\ref{fig:ed_tightbinding}.  A single exception at~$\gamma/\lambda = 9/11$, for which our analysis yields a zero octupole moment, disappears for $L{=}5$ and $L{=}7$, identifying a finite-size crossover.

These results demonstrate that although the three-dimensional realization is strongly finite in linear size, its momentum resolution and measurement window are sufficient for the topological phase classification used in the experiment.

\end{document}